\documentclass[12pt]{article} %***
\usepackage[sectionbib]{natbib}
\usepackage{array,epsfig,fancyheadings,rotating}
\usepackage[]{hyperref}  %<----modified by Ivan
\usepackage{sectsty, secdot}
\usepackage{float}
\usepackage{color,xcolor}
\sectionfont{\fontsize{12}{14pt plus.8pt minus .6pt}\selectfont}
\renewcommand{\theequation}{\thesection\arabic{equation}}
\subsectionfont{\fontsize{12}{14pt plus.8pt minus .6pt}\selectfont}
\usepackage{amsmath}
\usepackage{amssymb}
\usepackage{amsfonts}
\usepackage{multirow}
\usepackage{amsthm}

\newtheorem{theorem}{Theorem}

\newtheorem{corollary}{Corollary}

\theoremstyle{definition}

\newtheorem{example}{Example}
\newtheorem{remark}{Remark}

\newcommand{\bA}{{A}}

\newcommand{\bD}{{D}}

\newcommand{\bx}{x}

\newcommand{\um}{\underline{m}}

\newcommand{\bbT}{T}

\newcommand{\bI}{I}
\newcommand{\bbJ}{{\mathcal J}}

\newcommand{\bbS}{S}

\newcommand{\bX}{X}
\newcommand{\bU}{U}
\newcommand{\bV}{V}

\newcommand{\gS}{\Sigma}

\newcommand{\bPsi}{{\Psi}}

\newcommand{\bqa}{\begin{eqnarray}}
\newcommand{\eqa}{\end{eqnarray}}
\newcommand{\bqn}{\begin{eqnarray*}}
\newcommand{\eqn}{\end{eqnarray*}}
\newcommand{\be}{\begin{equation}}
\newcommand{\ee}{\end{equation}}
\newcommand{\non}{\nonumber\\}
\newcommand{\rE}{{\rm E}}

\newcommand{\md}{{\rm d}}

\allowdisplaybreaks[4]

\begin{document}

%%%%%%%%%%%%%%%%%%%%%%%%%%%%%%%%%%%%%%%%%%%%%%%%%%%%%%%%%%%%%%%%%%%%%%%%%%%%%%%%%%%%%%%%%%%%%%%%%%%%%%%%%%%%%%%%%%%%%%%%%%%%
%%%%%%%%%%%%%%%%%%%%%%%%%%%%%%%%%%%%%%%%%%%%%%%%%%%%%%%%%%%%%%%%%%%%%%%%%%%%%%%%%%%%%%%%%%%%%%%%%%%%%%%%%%%%%%%%%%%%%%%%%%%%

\renewcommand{\baselinestretch}{1.2}

%\markright{ \hbox{\footnotesize\rm Statistica Sinica
%{\footnotesize\bf 24} (201?), 000-000
%}\hfill\\[-13pt]
%\hbox{\footnotesize\rm
%\href{http://dx.doi.org/10.5705/ss.20??.???}{doi:http://dx.doi.org/10.5705/ss.20??.???}
%}\hfill }

\markboth{\hfill{\footnotesize\rm DANDAN JIANG} \hfill}
{\hfill {\footnotesize\rm TEST ON SPIKES IN A GENERALIZED SPIKED MODEL} \hfill}

\renewcommand{\thefootnote}{}
$\ $\par

%%%%%%%%%%%%%%%%%%%%%%%%%%%%%%%%%%%%%%%%%%%%%%%%%%%%%%%%%%%%%%%%%%%%%%%%%%%%%%%%%%%%%%%%%%%%%%%%%%%%%%%%%%%%%%%%%%%%%%%%%%%%

\fontsize{12}{14pt plus.8pt minus .6pt}\selectfont \vspace{0.8pc}
\centerline{\large\bf A UNIVERSAL TEST ON SPIKES IN A HIGH-}   
\vspace{2pt} 
\centerline{\large\bf  DIMENSIONAL GENERALIZED SPIKED }
\vspace{2pt} 
\centerline{\large\bf MODEL  AND ITS APPLICATIONS}
\vspace{.4cm} 
\centerline{Dandan Jiang } 
\vspace{.4cm} 
\centerline{\it School of Mathematics and Statistics}
\vspace{-1pt} 
\centerline{\it Xi'an Jiaotong University}
\vspace{-4pt} 
%\centerline{\it  Xianning West Road 28, Xi'an, Shannxi,  P.R. China.}
 \vspace{.55cm} \fontsize{9}{11.5pt plus.8pt minus.6pt}\selectfont

%%%%%%%%%%%%%%%%%%%%%%%%%%%%%%%%%%%%%%%%%%%%%%%%%%%%%%%%%%%%%%%%%%%%%%%%%%%%%%%%%%%%%%%%%%%%%%%%%%%%%%%%%%%%%%%%%%%%%%%%%%%%

\begin{quotation}
\noindent {\it Abstract:}
%{\bf Contents of the Abstract.}\\
This paper aims to test the number of spikes in a generalized spiked covariance matrix, the spiked eigenvalues of which may be extremely larger or smaller than the non-spiked ones. For a high-dimensional problem, we first propose a general test statistic and derive its central limit theorem by random matrix theory without a Gaussian population constraint. We then apply the result  to estimate the noise variance and test the equality of the smallest roots in generalized spiked models. Simulation studies showed that the proposed test method was correctly sized, and the power outcomes showed the robustness of our statistic to deviations from a Gaussian population. Moreover, our estimator of the noise variance resulted in much smaller mean absolute errors and mean squared errors than existing methods. In contrast to previously developed methods, we eliminated the strict conditions of {diagonal or block-wise diagonal form of the population covariance matrix}
and extend the work to a wider range without the assumption of normality. Thus, the proposed method is more suitable for real problems.

\vspace{9pt}
\noindent {\it Key words and phrases:}
Generalized spiked model; High-dimensional covariance matrix;
Testing the spikes; Central limit theorem.
\par
\end{quotation}\par

\def\thefigure{\arabic{figure}}
\def\thetable{\arabic{table}}

\renewcommand{\theequation}{\thesection.\arabic{equation}}

\fontsize{12}{14pt plus.8pt minus .6pt}\selectfont

%% main text
%%%%%%%%%%%%%%%%%%%%%%%%%%%%%%%%%%%%%%%%%%%%%%%%%%%
%      Introduction
%%%%%%%%%%%%%%%%%%%%%%%%%%%%%%%%%%%%%%%%%%%%%%%%%%%
\section{Introduction} \label{Int}
A generalized spiked model {from a population covariance matrix $\gS$ without  diagonal or block-wise diagonal assumption}  and a Gaussian population constraint
% without {diagonal or block-wise diagonal independence}  and a Gaussian population constraint 
is considered in this paper.  We let 
 $\bbT_p$ be a $p\times p$ deterministic matrix and 
$\gS=\bbT_p\bbT_p^*$ be a general population covariance matrix, which has 
the spikes $\alpha_1, \ldots, \alpha_K$ with multiplicity $m_k, k=1,\ldots,K$
arranged arbitrarily in groups among all the eigenvalues.
The condition $m_1+\cdots+m_K=M$ is satisfied, where $M$ is a fixed integer compared with the large dimension $p$. 
Furthermore, a few fixed eigenvalues (spikes) are 
allowed to be much larger or smaller than the majority of the eigenvalues.
This model is the so-called generalized spiked model.
The spiked model is 
closely related to principal component analysis (PCA) and factor analysis (FA), which are important and powerful tools for dimensionality reduction, data visualization, and feature extraction. 
Spiked models have also been
widely used in various scientific fields, such as factor models in economics and signal-plus-noise models in wireless communication. 
Whether the spikes have important effects on the identification of the key factors in a dataset must be determined.
Thus, the limiting behaviors of sample spiked eigenvalues and eigenvectors have attracted significant interest from researchers. 
One of the pioneering works was performed by \cite{Johnstone2001} when the population covariance was assumed to be 
 a high-dimensional identity matrix with fixed spikes. Under this simplified framework, \cite{Baik2005}, \cite{BaikSilverstein2006}, \cite{Paul2007}, and \cite{BaiYao2008} examined the limiting results
 of sample spiked eigenvalues in different aspects. 
 As a further step, \cite{BaiYao2012}, \cite{FanWang2015}, \cite{CaiHanPan2017}, and \cite{JiangBai2021a, JiangBai2021b} extended the structure of the population covariance to a more general form and investigated the asymptotic distributions of the sample spiked eigenvalues in high-dimensional settings. 
 
 However, there are relatively fewer studies on testing the number of the spikes for large dimensionality $p$, which is generally a fundamental step to reconstruct the structure of the population covariance. 
 Most of the related work was performed to determine the number of spikes 
 in a high-dimensional setting based on random matrix theory, such as in the work of 
 \cite{KritchmanNadler2008} and \cite{PassemierYao2012}. 
 In contrast, 
\cite{Onatski2009}, \cite{Passemieretal2017}, and \cite{JO2019} performed work relevant to tests of the spikes. 
\cite{JO2019} focused on testing the existence of spikes.
 \cite{Passemieretal2017} derived a goodness-of-fit test for a high-dimensional principal component model and determined the number of the principal components.
 \cite{Onatski2009} developed a test of the number of factors in large factor models with a white noise assumption. 
These previous approaches were limited in various ways, such as requiring {diagonal or block-wise diagonal  form of the population covariance matrix,}  a Gaussian population assumption or  only including extremely large spikes but not extremely small ones.

To relax these restrictions,
we recall the generalized spiked model described {at the beginning} and provide a corrected pseudo-likelihood
ratio test on the number of spikes for this model. The proposed test is universal for all population assumptions and the general form of the spiked covariance matrix. This test was applied to 
estimate the noise variance and testing equality of the smallest roots in the generalized spiked model.
In comparison with previous studies, our proposed test method has the following advantages. First, we extended the population covariance matrix to a general non-negative definite matrix and removed {its diagonal or block-wise diagonal  assumption}. Second, the asymptotic distribution of the proposed test statistic was established 
in a high-dimensional setting without the Gaussian population assumption.
Moreover, under this setting, the spikes are allowed to be significantly larger or smaller than the non-spiked eigenvalues, occurring in several bulks.
Overall, our assumptions are more practical than those imposed in previous work.

An outline of this paper is as follows. In Section~\ref{Sec2}, the problem is described in a generalized setting, and
a universal test  on the spikes is proposed  {without  the constraints of  the  Gaussian and  diagonal or block-wise diagonal assumptions}. 
Section~{\ref{Sec3} and}~\ref{Sec4} gives two important applications: one is to estimate the noise variance; the other is to test the equality of the smallest roots in generalized spiked models. Simulations are  also conducted for each result to evaluate our work comparing with the existing works. Then,  based on the above conclusions, we also analyze the two sets of real data, and give the corresponding statistical inference in Section~\ref{Sec5}.
Finally, we draw a conclusion in the Section~\ref{Sec6}. Detailed proofs are all provided in the Appendix. 

% 
%%%%%%%%%%%%%%%%%%%%%%%%%%%%%%%%%%%%%%%%%%%%%%%%%%%
%    Main result
%%%%%%%%%%%%%%%%%%%%%%%%%%%%%%%%%%%%%%%%%%%%%%%%%%%
\section{ Test on spikes in 
high-dimensional generalized spiked model.
} \label{Sec2}

We consider the generalized spiked model mentioned in the Introduction, which was first proposed by \cite{JiangBai2021a},
and define the singular value decomposition of $\bbT_p$ as
 \be \bbT_p= \bV\left(
\begin{array}{cc}
 \bD_1^{1 \over 2} & 0 \\
 0 & \bD_2^{1 \over 2} 
\end{array}
\right)\bU^{*},
\label{UDU}
\ee
where $\bU$ and $\bV$ are unitary matrices, $\bD_1$ is a diagonal matrix of the $M$ spiked eigenvalues,
 and $\bD_2$ is the diagonal matrix of the non-spiked eigenvalues with bounded components. We define $\bU_1$ and  $\bU_2$ as the 
  first $M$ and the last $p-M$ columns of matrix $\bU$ defined in Eq. (\ref{UDU}), respectively.
 
We assume that the double array $\{x_{ij},i,j = 1,2,...\} $ consists of ${\rm i.i.d.}$ random variables with mean 0 and variance 1. 
Furthermore,   $\rE (x_{ij})=0$ and $\rE (x_{ij}^2)=0$ for the complex case.
Thus, 
\begin{equation}
\bbT_p\bX=(\bbT_p\bx_1,\cdots,\bbT_p\bx_{n}) 
\label{TpX}
\end{equation}
can be seen as a random sample from the population with general population covariance matrix $\gS$,
 where $\bx_j=(x_{1j}, \cdots, x_{pj})', 1\leq j \leq n$.
The corresponding sample covariance matrix of observations $\bbT_p\bX$ is 
\begin{equation}
\bbS=\bbT_p\left(\frac{1}{n}\bX\bX^*\right)\bbT_p^*,
\label{S}
\end{equation}
which is the generalized spiked sample covariance matrix.

To test the number of the spikes, the following hypothesis is used:
\begin{equation}
\mathcal{H}_0: M=M_0 \quad \mbox{v.s.} \quad M \neq M_0,
\label{H0}
\end{equation}
{where $M_0$  is a given non-negative integer, such that  \eqref{H0} is to test whether the true number of the spikes is $M_0$. We consider the hypothesis testing  in \eqref{H0}   under 
 the high-dimensional setting,  which assumes} that $p/n = c_n \to c >0$ and both $n$ and $p$ tend to infinity simultaneously.

We define the empirical spectral distribution (ESD) of $\gS$ 
as $H_n(t)$, which tends to a proper probability measure $H(t)$ as $p \rightarrow \infty$. 
We let 
{
\begin{equation}
\bbJ_k=\{ j_k+1,\ldots, j_k+m_k\} \label{Jk}
\end{equation}
}
denote the set of ranks of the $m_k$-ple eigenvalue $\alpha_k$ in the descending %array of all the
 population eigenvalues, where $\alpha_k$ is out of the support of $H(t)$.
Moreover, we define  $\{l_{j}(\bbS), j \in \bbJ_k\}$, $k=1,\ldots,K$ are the 
 associated sample eigenvalues of the matrix $\bbS$, which are simply denoted as $l_j$ henceforth. 
By Proposition~2.1 of \cite{JiangBai2021a}, 
for each population spiked eigenvalue 
$\alpha_k$ with multiplicity $m_k$ satisfying the 
 the separation condition
$\min_{i\ne k}|\alpha_k/\alpha_i-1|>d$,  
where $d$ is some positive constant,
we have ${l_{j}(\bbS)}/{\phi_k}-1 \to 0, a.s.$ for all $j \in \bbJ_k$ and the function $\phi(x)=x\{1+c\int{t}/{(x-t)} \md H(t)\}$.
This conclusion holds 
 under the bounded 4th-moment assumption. 
 However, based on the truncation procedures of \cite{JiangBai2021a}, the convergence in Proposition~2.1 still holds in probability without the bounded 4th-moment assumption but the one of the tail probability is satisfied,
  i.e.
 \be
\lim\limits_{\tau \rightarrow \infty}\tau^4 {\rm P}\left(|x_{ij}| >\tau \right)=0.
\label{tail}
 \ee

Inspired by this result, we propose a test statistic for (\ref{H0}) and derive its asymptotic distribution. 
Recall the 
likelihood ratio test statistic for (\ref{H0}) in the probabilistic principal component analysis model 
$\gS=\mbox{diag}(a_1,\cdots,a_M,0\cdots,0)+\sigma^2\bI$ is expressed by 
\be
L =\left[\frac{1}{p-M}\sum\limits_{i=M+1}^p l_i\left( \prod_{i=M+1}^p l_i\right)^{-\frac{1}{p-M}}\right]^{-\frac{(p-M)n}{2}}\label{Lstat}
\ee
in classical statistical theory.
The test statistic $-2\log L$ 
 relies mainly on the partial linear spectral statistic involved with the non-spiked 
 eigenvalues, such as $\sum_{i=M+1}^p l_i$ and $\sum_{i=M+1}^p \log l_i$. 
Following 
Anderson and Rubin (1956), this type of statistic is also used in the goodness-of-fit test for
the probabilistic principal component analysis model.
Therefore, for the generalized spiked model, we propose the statistic 
\be
\sum\limits_{j=1}^p f(l_j)-\sum\limits_{j \in \bbJ_k, k=1}^Kf( l_j),\label{newtest}
\ee
where 
{$\bbJ_k$ is defined as \eqref{Jk}, }
$f \in \mathcal{A}$,  and  $\mathcal{A}$ is a set of analytic functions defined on an open set of the complex plane
 including the whole supporting set of the limiting spectral distribution (LSD) $H(t)$. 
We define  $F^{c,H}$ as  the LSD of the sample matrix $\bbS$, and $F^{c_n,H_n}$ is the analogue of $F^{c,H}$  with the 
$c, H$ substituted by $c_n, H_n$, respectively. Furthermore, the 
 $\underline{m}(z)\equiv m_{\underline{F}^{c,H}}(z)$ is defined as the
Stieltjes transform of $\underline{F}^{c,H}\equiv (1-c)I_{[0,
  \infty)}+cF^{c,H}$. To obtain the asymptotic distribution of the 
test statistic (\ref{newtest}), some assumptions are also given as below:
\begin{description}
\item[Assumption~(a): ]  The tail probability
(\ref{tail}) is satisfied
 and
 $p/n = c_n \to c >0$ as  both $n, p \to \infty$;\\[-1cm]
  \item[Assumption~(b): ]  Assume  that
  $
\lim \sum_{t=1}^p |u_{ti}|^4\rE\{|x_{11}|^4I(|x_{11}|\le \sqrt{n})-2-q\} < \infty$,
 where $q=1$ for the real case, $q=0$ for the complex case,  $I ( \cdot )$ is 
the indicator function, 
and 
 $u_i=(u_{1i}, \ldots, u_{pi})'$ is the $i$th column of the matrix $\bU_1$.\\[-1cm]
  \item[Assumption~(b*): ]   Suppose that  \be
 \max\limits_{1\leq t \leq p,1 \leq i \leq M} |u_{ti}|^2 \rE\big\{|x_{11}|^4I(|x_{11}|<\sqrt{n})-2-q\big\}\rightarrow 0.  % ~~ t=1,\ldots,p; i=1,\ldots,M.
 \label{CondU1}
 \ee
\end{description}

 Thus, the central limit theorem (CLT) for the test statistic (\ref{newtest}) is established as follows,
 and is proven in the Appendix.
 \begin{theorem}\label{Th1}
 For the testing problem (\ref{H0}), suppose that Assumption~(a) and (b)  [or   (b*)] hold simultaneously.
Then
the asymptotic distribution of the 
test statistic (\ref{newtest}) is as follows:
 \be
T_{f,H}= \nu_{f,H}^{-\frac{1}{2}}\left\{\sum\limits_{j=1}^p f(l_j)-\sum\limits_{j \in \bbJ_k, k=1}^Kf( l_j)
 -{b_{f,H_n}}-\mu_{f,H}\right\}\Rightarrow \mathcal{N} \left( 0,
 1\right), \label{CLT1}
 \ee
 where
 \begin{align}
{b_{f,H_n}}&={p \int f(t) {\rm d} F^{c_n,H_n}(t)-\sum\limits_{k=1}^Km_kf\left(\alpha_k+
 c\alpha_k\int\frac{t}{\alpha_k-t} \md H(t)\right)},\non
  \mu_{f,H}&=\!-\frac{q}{2\pi i} \!\oint \!f(z) \frac{\!c \int \underline{m}^3(z)t^2\{1+t\underline{m}(z)\}^{-3} \md H(t)}{\left[1\!-\!c\!\int \underline{m}^2(z)t^2 \{1+t\underline{m}(z)\}^{-2}\md H(t)\right ]^2 } \md z \non
&-\frac{\beta c }{2 \pi i} \oint f(z) \frac{ \underline{m}^3(z)\cdot \int t\{1+t\underline{m}(z)\}^{-1} \md H(t)\cdot \int \{1+t\underline{m}(z)\}^{-2} \md H(t)}
{1\!-\!c\!\int \underline{m}^2(z)t^2 \{1+t\underline{m}(z)\}^{-2}\md H(t)} \md z,\non
\nu_{f,H}&=-\frac{q+1}{4\pi^2}\oint\oint\frac{f(z_1)f(z_2)}{\{\underline{m}(z_1)-\underline{m}(z_2)\}^2}
\md\underline{m}(z_1)\md \underline{m}(z_2) \non
&-\frac{\beta c}{4\pi^2}\oint\oint f(z_1)f(z_2)\int \frac{t \md H(t)}{\{1+t\underline{m}(z_1)\}^2}\! \int \frac{t \md H(t)}{\{1+t\underline{m}(z_2)\}^2} \md\underline{m}(z_1)\md \underline{m}(z_2).\label{04var2}
 \end{align}
 Here $\beta=\lim \sum_{t=1}^p |u_{ti}|^4\rE\{|x_{11}|^4I(|x_{11}|\le \sqrt{n})-2-q\}$ if Assumption~(b) is met, and 
 $\beta=0$ if Assumption~(b*)  holds instead. 
The contours all contain the support of $F^{c,H}$ and 
  are non-overlapping in (\ref{04var2}).
   \label{Th1}
 \end{theorem}

  \begin{remark}
  Note that the CLT given by Theorem~\ref{Th1} depends on the number $m_k$,
  the values of the population spikes $\alpha_k$, and the LSD $H(t)$. However, these parameters are unknown in real data analysis and need to be estimated.
  Some helpful estimation methods can be found in various references
   (\cite{Lietal2013, JiangBai2021a, Baoetal2019, Zhengetal2021}).
  \end{remark}
For $H(t)=\delta_{\{1, +\infty\}}$, we select $f(x)=x$ and $f(x)=\log x$ as two examples and present the details of the computations in  the supplementary material.

\begin{example}\label{eg:1}
If $f(x)=x$ and $H(t)=\delta_{\{1, +\infty\}}$, the statistic in (\ref{CLT1}) simplifies to
\[
T_{x,1}= \nu_{x,1}^{-\frac{1}{2}}\left\{\sum\limits_{j=1}^p l_j-\sum\limits_{j \in \bbJ_k, k=1}^K l_j
 -b_{x,1}-\mu_{x,1}\right\}\Rightarrow \mathcal{N} \left( 0,
 1\right), 
\]
 where $b_{x,1}\!=\!(p-M)\!-\!\sum_{i=1}^K \!{m_kc\alpha_k}/{(\alpha_k-1)}$, $\mu_{x,1}=0$, and $\nu_{x,1}\!=\!(q+1+\beta)c$.
\end{example}

\begin{example}\label{eg:2}
If $f(x)=\log x$ and $H(t)=\delta_{\{1, +\infty\}}$, the statistic in  (\ref{CLT1}) is given as 
\[
T_{\log,1}= \nu_{\log,1}^{-\frac{1}{2}}\left\{\sum\limits_{j=1}^p\log ( l_j)-\sum\limits_{j \in \bbJ_k, k=1}^K \log
( l_j)
 -b_{\log,1}-\mu_{\log,1}\right\}\Rightarrow \mathcal{N} \left( 0,
 1\right), 
\]
 where
\begin{align*}
b_{\log,1}&=p\left\{\frac{(c-1)}{c}\log(1-c)-1\right\}-\sum\limits_{i=1}^K m_k \log (1+\frac{c}{ \alpha_k-1}), \non
\mu_{\log,1}&=\frac{q}{2}\log (1-c)-\frac{1}{2}\beta c, \quad \nu_{\log,1}=-(q+1)\log (1-c)+\beta c.
\end{align*}
\end{example}

\subsection{Monte Carlo experiments}\label{Sec21}

To illustrate the effectiveness of the proposed CLT with simulations, we first provide the following two models: 
\begin{description}
\item[ Model~1:]  Assume that $\gS=\Lambda$, where $\Lambda$ is assumed to be an identity matrix with a finite-rank perturbation.
Its spikes 
are $(25,16,16,0.2,0.2,0.1)$ with the multiplicities $(1,2,2,1)$ in descending order. 
\item[Model~2:] Assume that $\gS=\bU_0 \Lambda \bU_0^*$, where $\Lambda$ is defined in Model 1 and 
$\bU_0$ is the matrix composed of the eigenvectors
 of a $p\times p$
matrix with the entries being independently sampled from $N(0,1)$. Thus,  {the diagonal  assumption of $\gS$} is relaxed.
\end{description}

Furthermore, to show that the conclusion is widely applicable and free of the population assumption, 
the  Gaussian and Gamma populations are considered 
for each model:
\begin{description}
\item[ Gaussian Assumption:] $\{x_{ij}\}$ are ${\rm i.i.d.}$ samples from a standard Gaussian population. 
 \item[ Gamma Assumption:] $\{x_{ij}\}$ are i.i.d. from the population distribution $Gamma (4,0.5)-2$. 
\end{description} 

{
To further demonstrate the proposed CLT  is valid for a distribution with infinite fourth moments under the Assumption~(b*), we generate i.i.d. samples $x_{ij}$  from
$2^{-1/2}t(4)$ population distribution under the setting of Model~2, while the fourth moments of
$x_{ij}$ are infinite.}

\begin{figure}[t!]
\includegraphics[width = .3\textwidth] {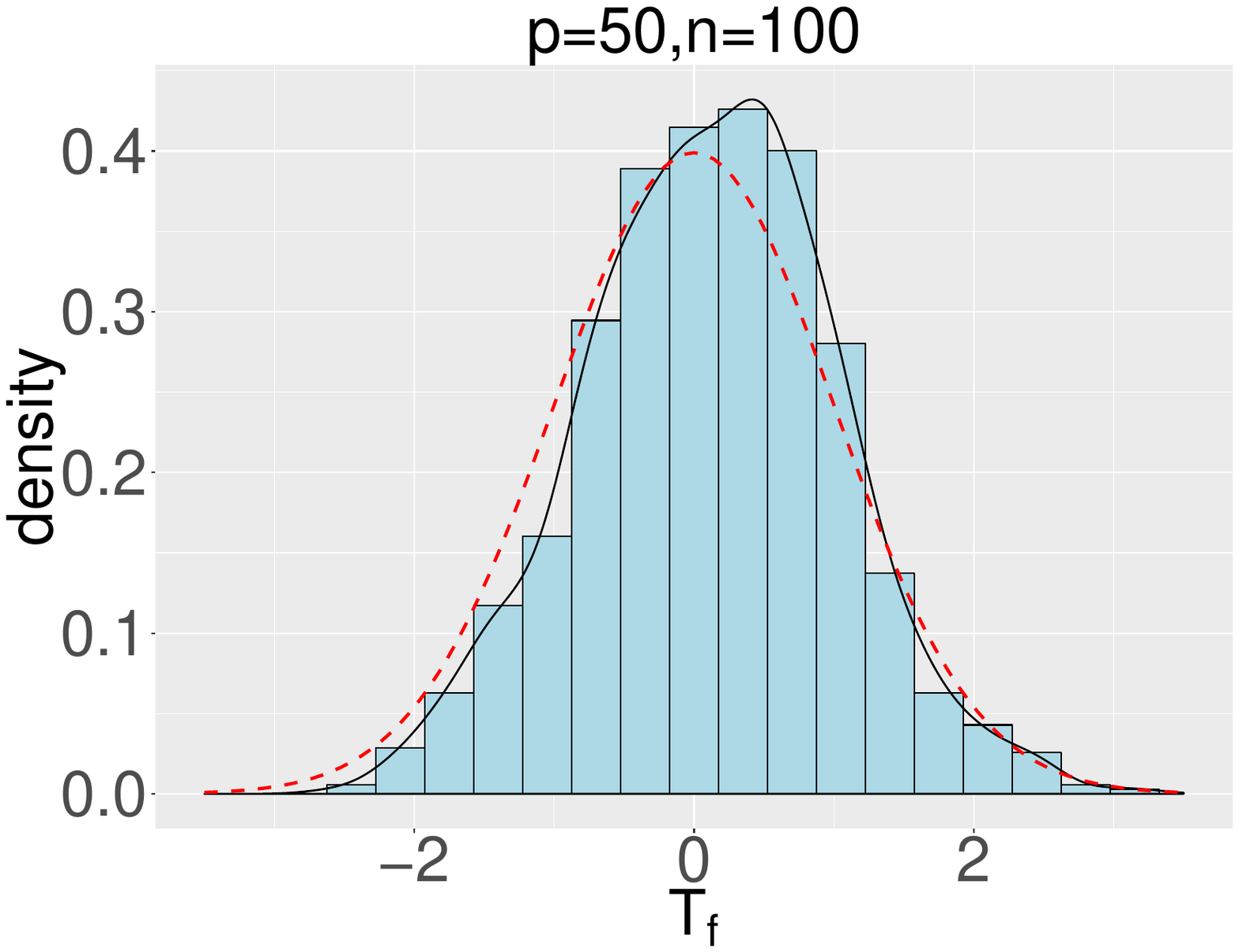}
\quad\includegraphics[width = .3\textwidth] {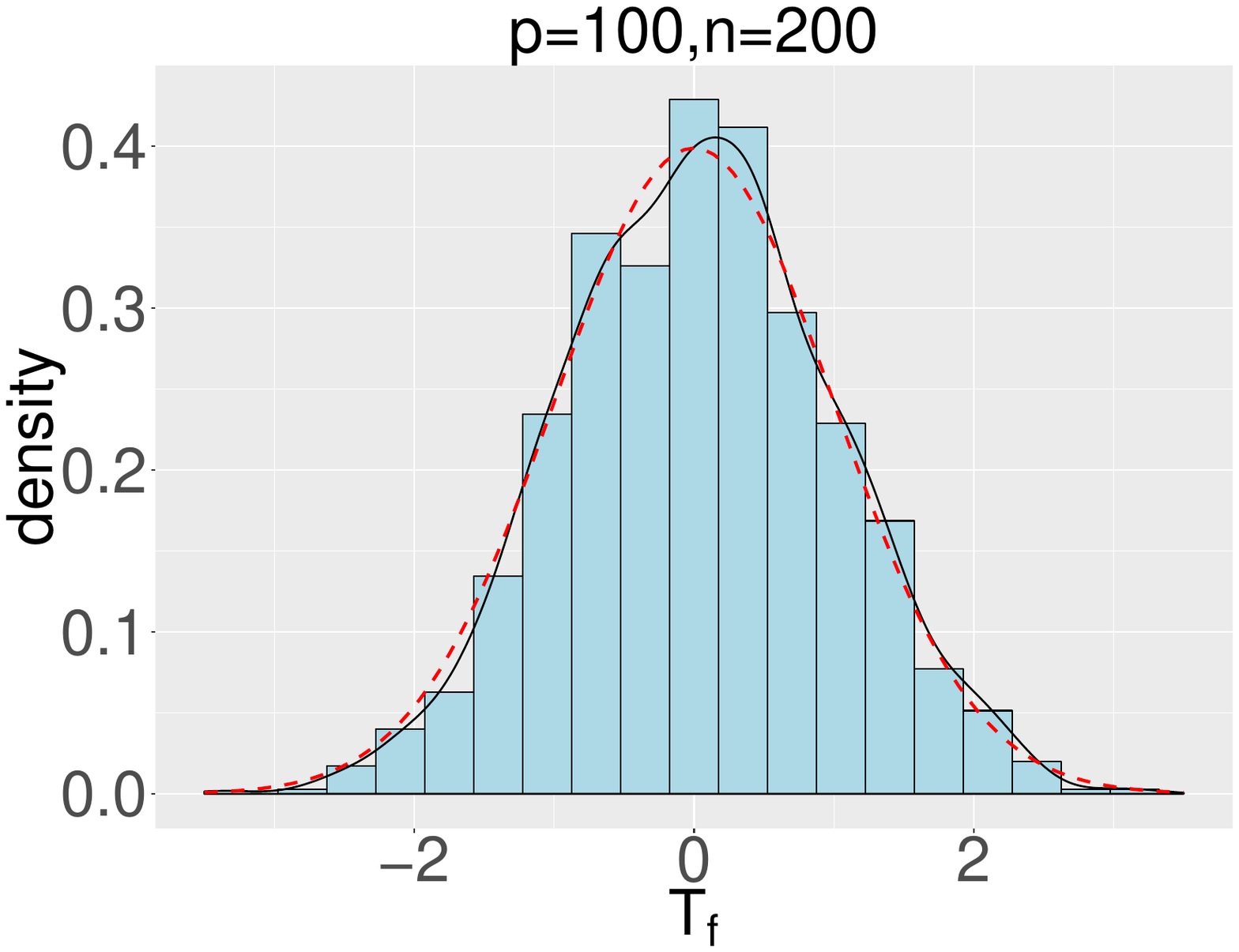}\quad
\includegraphics[width = .3\textwidth] {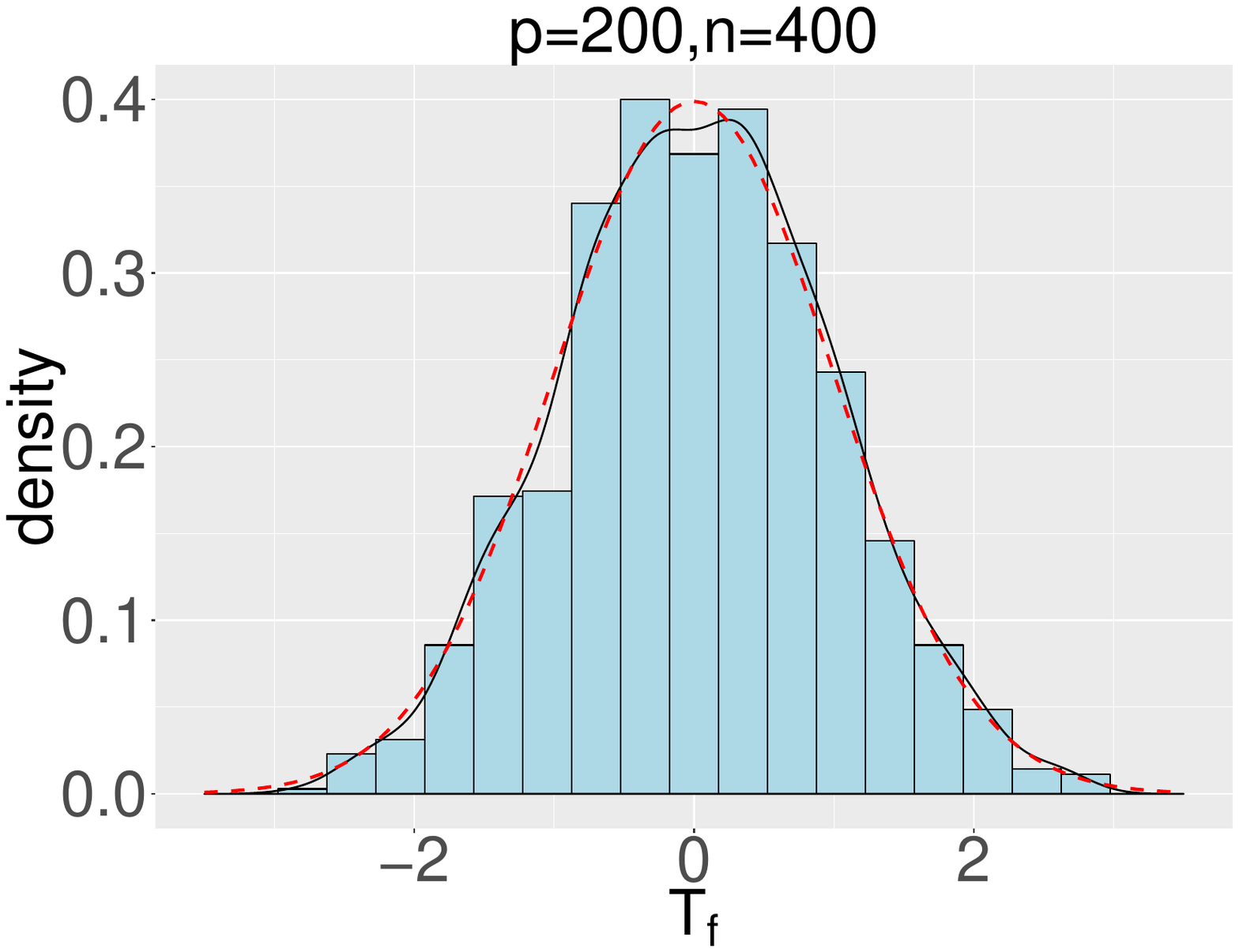}\\[-.5mm]
\includegraphics[width = .3\textwidth] {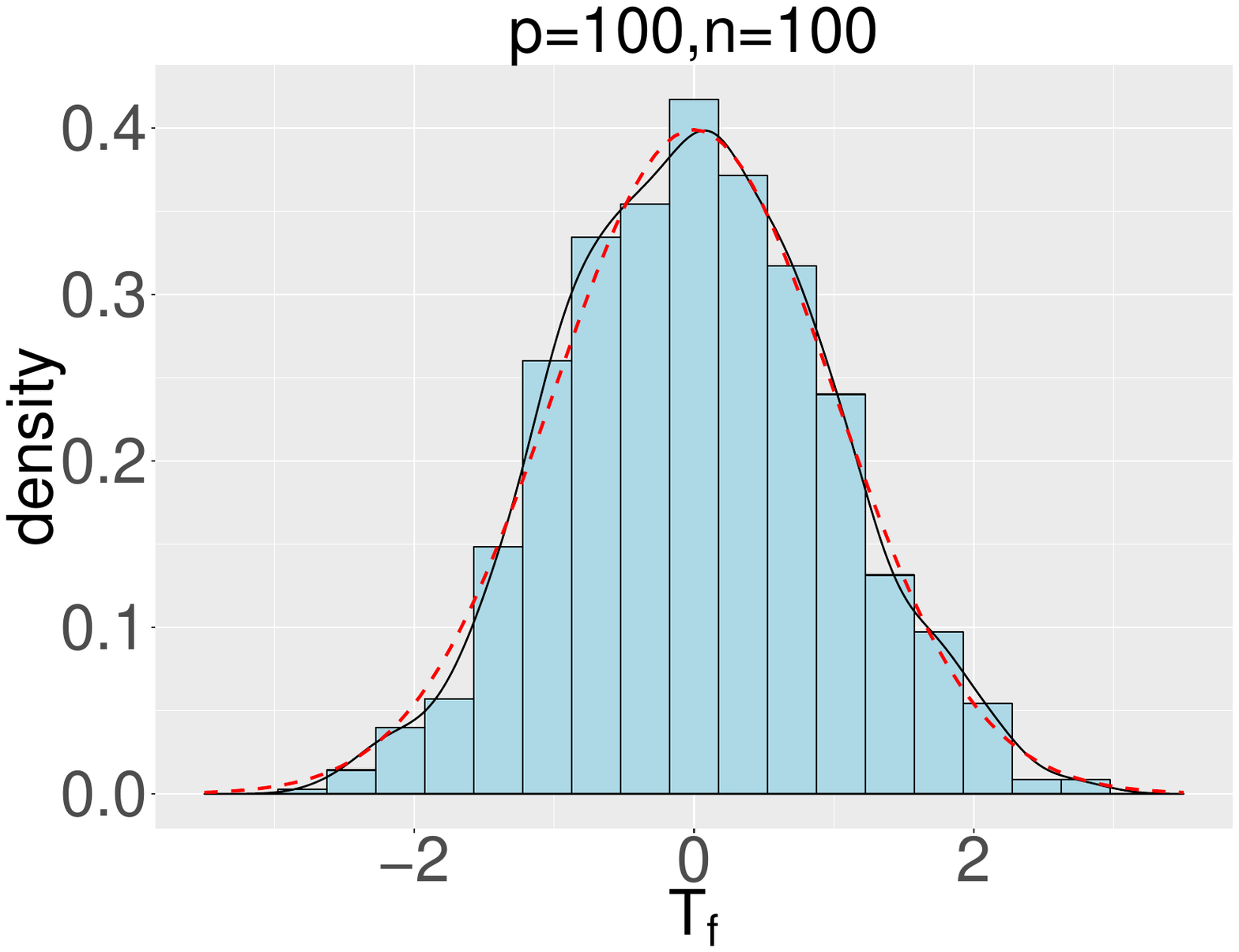}
\quad\includegraphics[width = .3\textwidth] {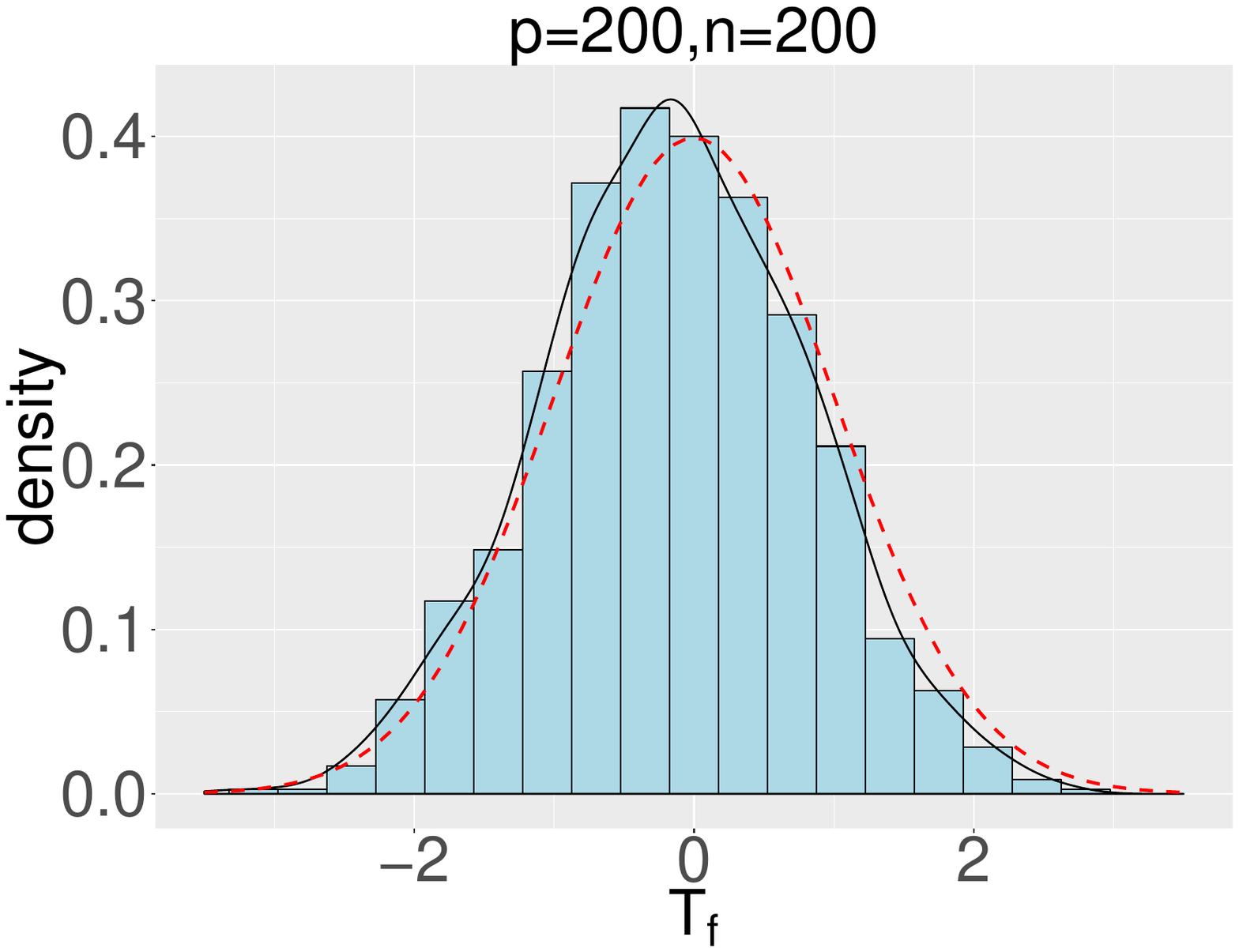}\quad
\includegraphics[width = .3\textwidth] {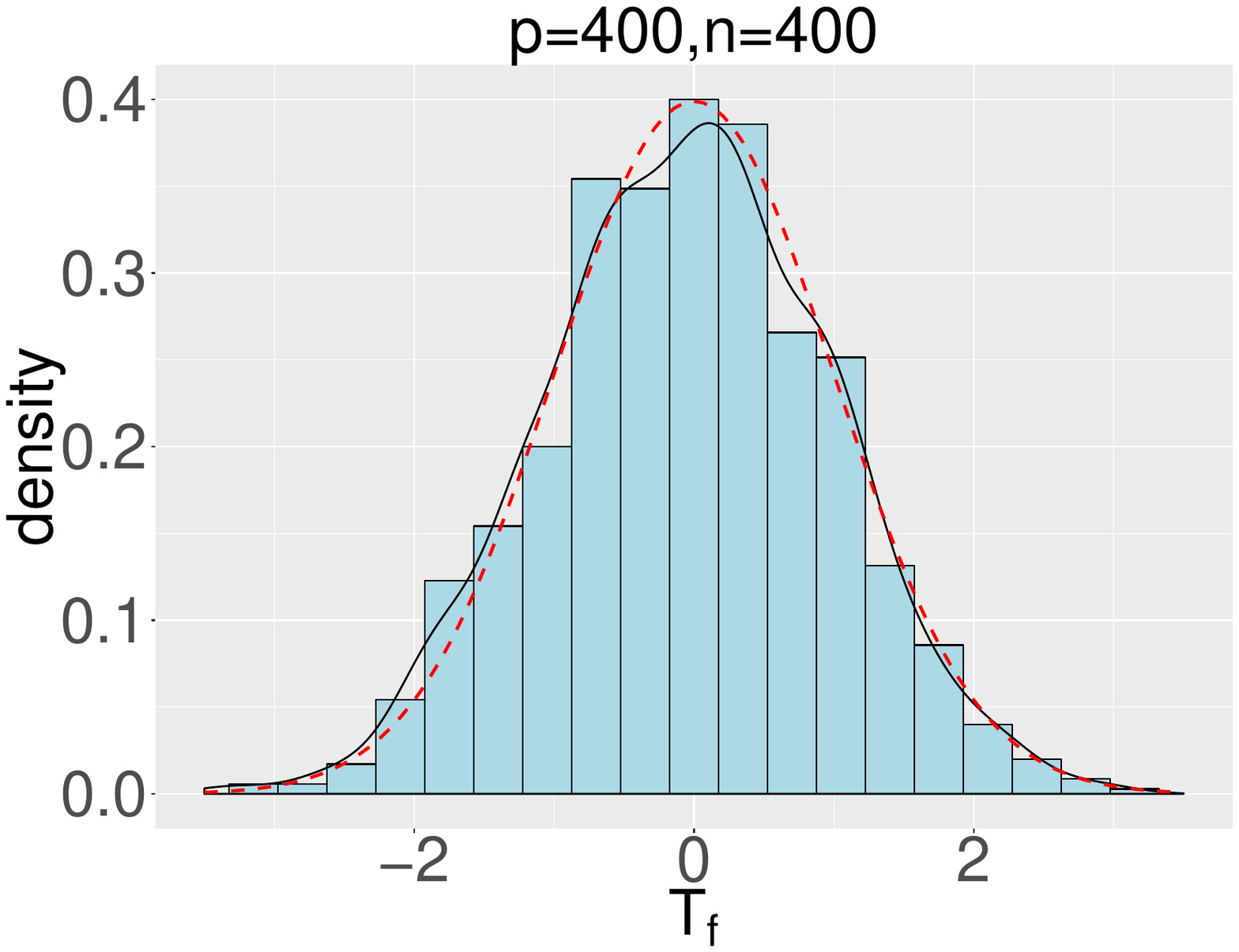}\\[-.5mm]
\includegraphics[width = .3\textwidth] {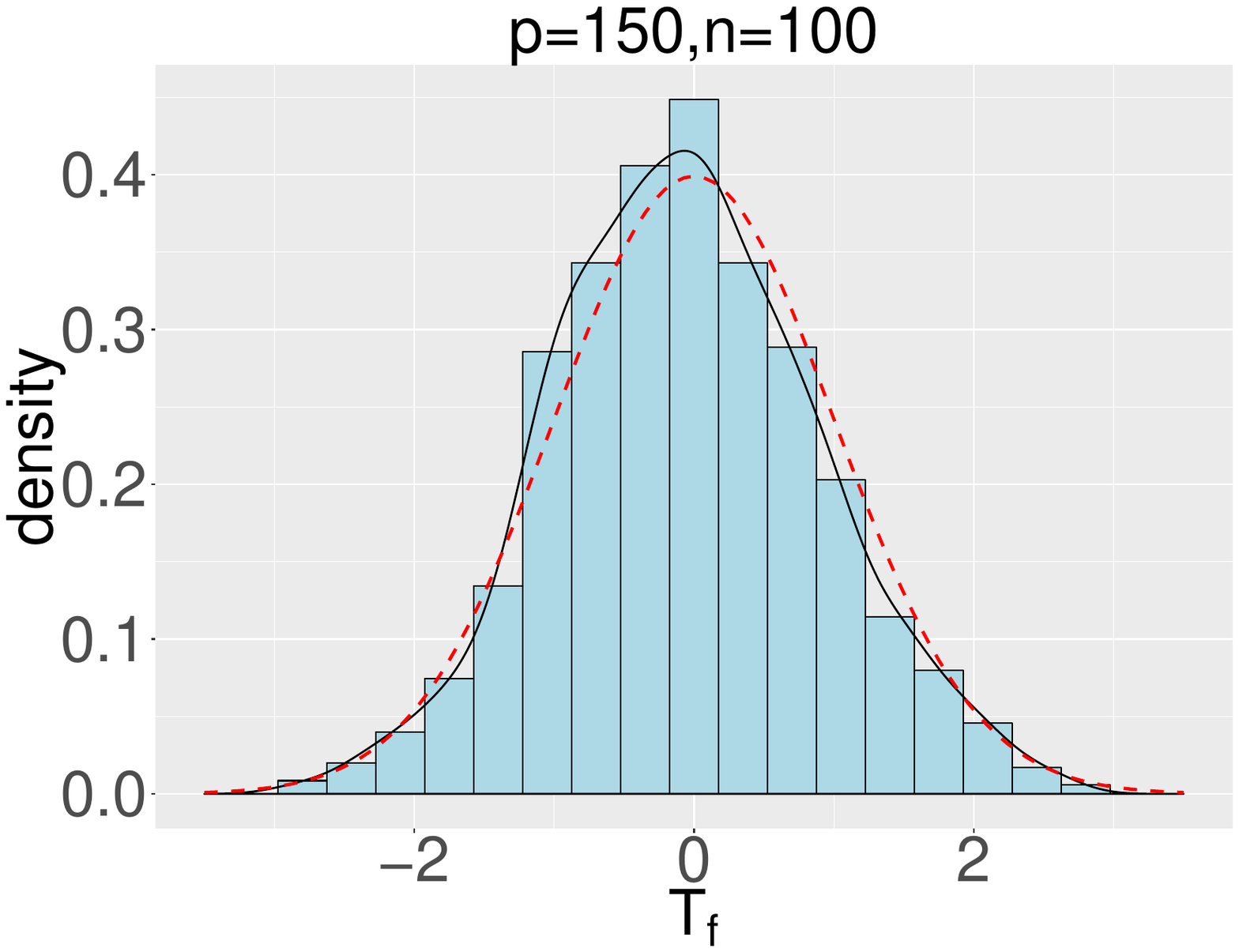}
\quad\includegraphics[width = .3\textwidth] {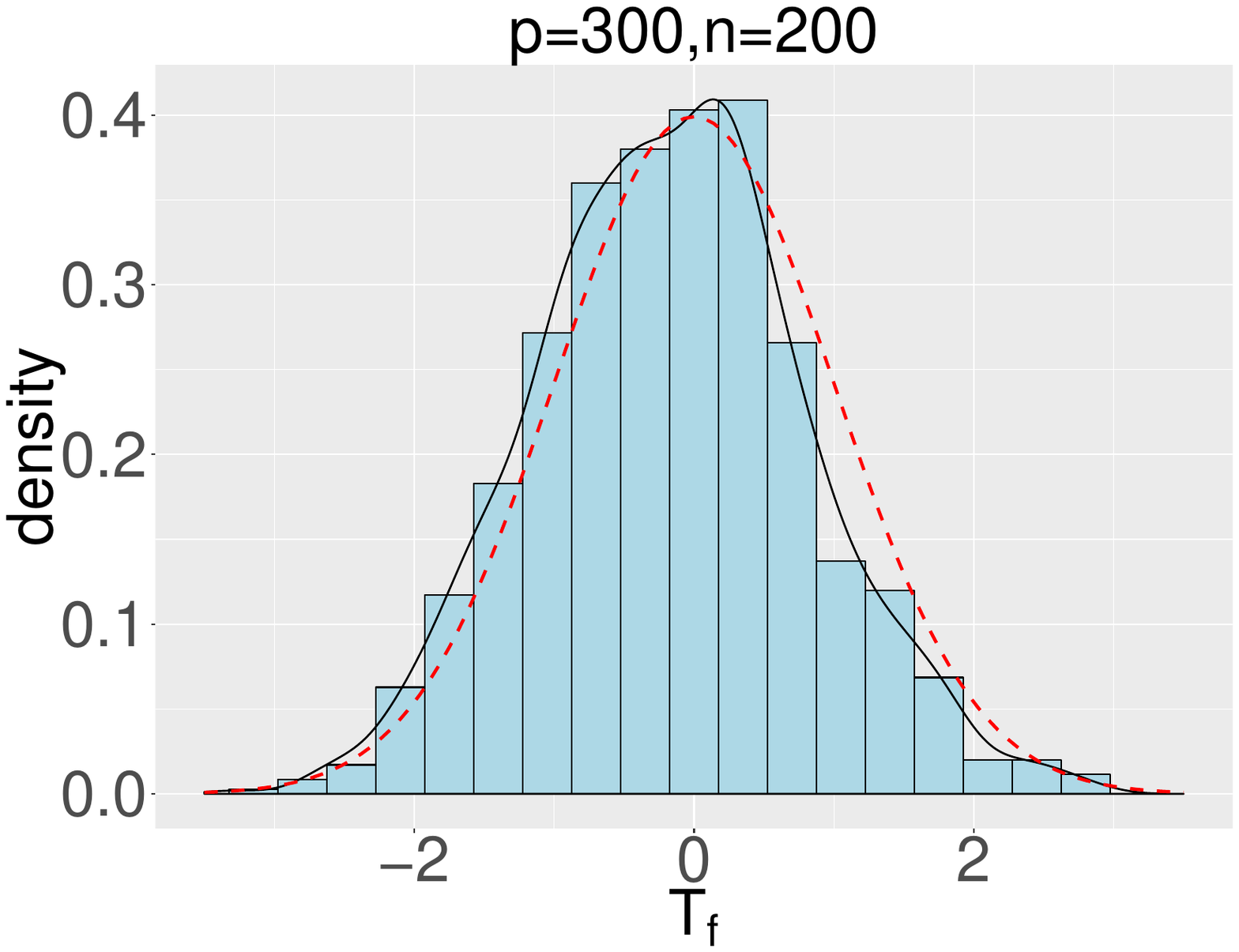}\quad
\includegraphics[width = .3\textwidth] {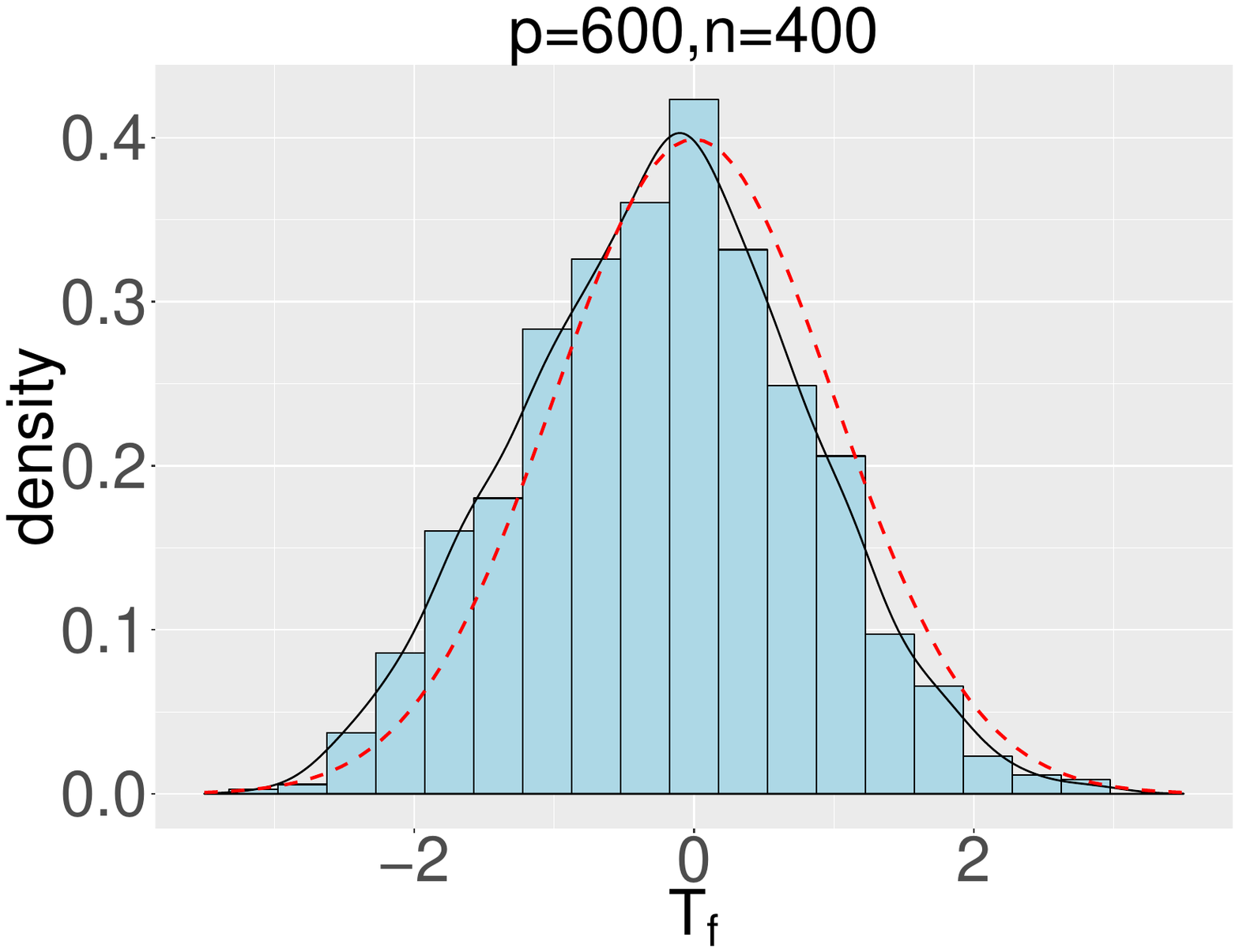}
\caption{ {\rm {Model~2}} under Gamma assumption. 
 }\label{fig:4}
\end{figure}

For the sake of simplicity, we selected the function $f(x)=x$ and the LSD $H(t)=\delta_{\{1, +\infty\}}$.
For each case, the 
sample size was set to $n=100$, $200$, and $400$ and the values of $p/n=0.5$, $1$, and $1.5$, respectively.
We report 
the empirical probability of rejecting the null hypothesis (\ref{H0}), $\mathcal{H}_0: M=M_0$, 
with 1000 replicates in Table~\ref{tab:11} and Table~\ref{tab:12}. We only list the results for Model~2   here, and similar results for Model~1 are listed in the supplementary material for brevity.
Moreover, the empirical distributions of the proposed test statistic when $M_0$ was equal to the true value of $M$ are also plotted.
Figure~{\ref{fig:4}} shows the performance of our proposed method under Model~2 with a Gamma population assumption, and the figures for the other cases are also shown 
in the supplementary material. 

The simulation study illustrated that the proposed test on the number of the spikes
provided good sizes for both Gaussian and non-Gaussian, diagonal and off-diagonal populations when the null hypothesis was true.
Moreover, the further the alternative hypothesis was from the null hypothesis, the higher the power was.
Based on the above results, we inferred that 
{the given number  $M_0$ being tested in \eqref{H0}  that matches the true value of $M$ usually occurs at the inflection point of the empirical sizes.} This corresponds to the first local minimum value of the empirical sizes.

\begin{table}[t!] 
\caption{Empirical probability of rejecting the null hypothesis (\ref{H0}) for Model~2 under Gaussian assumption, when the true value of $M$ is 6.}
\label{tab:11}\par
\resizebox{\linewidth}{!}{\begin{tabular}{|lccccccc|} \hline %***5truept
% \multicolumn{8}{|c|}{Model~2 under Gaussian assumption} \\ \hline
{Values of $M_0$} & 1&2&3&4&5&6&7 \\[3pt] \hline
$p=50~;~ n=100$ & 1     & 1    &0.649 & 0.319 & 0.096 & \bf 0.048 &0.068 \\[-.5mm]
$p=100;~ n=200$ & 1     & 1    &0.691 & 0.366 & 0.116 & \bf 0.049 &0.115 \\[-.5mm]
$p=200;~ n=400$ & 1     & 1    &0.698 & 0.389 & 0.129 & \bf 0.055 &0.150 
\\ \hline
$p=100;~ n=100$ & 1     & 1    &0.383 & 0.195 & 0.085 & \bf 0.044 &0.125 \\[-.5mm]
$p=200;~ n=200$ & 1     & 1    &0.411 & 0.223 & 0.099 & \bf 0.042 &0.174 \\[-.5mm]
$p=400;~ n=400$ & 1     & 1    &0.403 & 0.231 & 0.121 & \bf 0.052 &0.213 \\ \hline
$p=150;~ n=100$ & 1     & 0.998 &0.263 & 0.151 & 0.078 & \bf 0.039 &0.174 \\[-.5mm]
$p=300;~ n=200$ & 1     & 1     &0.286 & 0.178 & 0.102 & \bf 0.040 &0.244 \\[-.5mm]
$p=600;~ n=400$ & 1     & 1     &0.301 & 0.199 & 0.124 & \bf 0.054 &0.263 \\ \hline
\end{tabular}}
\end{table}

\begin{table}[t!] 
\caption{Empirical probability of rejecting the null hypothesis (\ref{H0}) for Model~2 under Gamma assumption, when the true value of $M$ is 6.}
\label{tab:12}\par
\resizebox{\linewidth}{!}{\begin{tabular}{|lccccccc|} \hline %***5truept
%\multicolumn{8}{|c|}{Model~2 under Gamma assumption}\\ \hline
{Values of $M_0$} & 1&2&3&4&5&6&7\\ \hline
$p=50~;~ n=100$ & 1     &1     &0.403 & 0.190 & 0.071 & \bf 0.038 &0.051 \\[-.5mm]
$p=100;~ n=200$ & 1     & 1    &0.446 & 0.235 & 0.085 & \bf 0.046 &0.086 \\[-.5mm]
$p=200;~ n=400$ & 1     & 1    &0.456 & 0.236 & 0.101 & \bf 0.043 &0.112 \\ \hline
$p=100;~ n=100$ & 1     &0.996 &0.242 & 0.115 & 0.060 & \bf 0.040 &0.086 \\[-.5mm]
$p=200;~ n=200$ & 1     & 1    &0.282 & 0.155 & 0.082 & \bf 0.044 &0.127 \\[-.5mm]
$p=400;~ n=400$ & 1     & 1    &0.292 & 0.164 & 0.084 & \bf 0.049 &0.144 \\ \hline
$p=150;~ n=100$ & 1     & 0.980 &0.137 & 0.078 & 0.055 & \bf 0.043 &0.094 \\[-.5mm]
$p=300;~ n=200$ & 1     &0.991  &0.177 & 0.122 & 0.077 & \bf 0.044 &0.148 \\[-.5mm]
$p=600;~ n=400$ & 1     & 0.998 &0.214 & 0.125 & 0.091 & \bf 0.052 &0.181 \\ \hline
\end{tabular}}
\end{table}

%%%%%%%%%%%%%%%%%%%%%%%%%%%%%%%%%%%%%%%%%%%%%%%%%%%
%     App
%%%%%%%%%%%%%%%%%%%%%%%%%%%%%%%%%%%%%%%%%%%%%%%%%%%

\section{Application to estimation of noise variance in a generalized spiked model}\label{Sec3}
We suppose the spiked model described in Section~\ref{Sec2} has the following structure
\be
\gS=\bA\bA'+\bPsi,\label{51spikedmodel}
\ee
where $\bA$ is a $p\times M$ matrix, the eigenvalues of $\bA'\bA$
are $M$ distinct elements, and the eigenvalues of $\bPsi$
are several bulks of the general population eigenvalues. Thus, the
spectrum of $\gS$ is denoted as 
\be
\sigma^2({\underbrace{\tilde\alpha_1, \cdots,\tilde\alpha_1}_{m_1}}, \cdots,\underbrace{\tilde\alpha_k,\cdots, \tilde\alpha_k}_{m_k}, \cdots,
\underbrace{r_1, \cdots,r_1,\cdots, r_s,\cdots, r_s }_{p-M}),\label{Spec}
\ee
where $m_1+\cdots+m_k=M$,
and both $M$ and $s$ are fixed small numbers.

Since it is a relatively common phenomenon in practical applications that the general population eigenvalues are divided into several bulks with fixed distinct spikes, we assume that the LSD of $\gS$ excluding the spikes, denoted as $H(t)$, follows a probability distribution, which takes the value $r_i\sigma^2$ with probability $\omega_i$, $i=1,\cdots,s$, where $\omega_1+\cdots+\omega_s=1$.

In traditional statistical theory, the statistic 
\be
\hat\sigma^2= \frac{1}{(p-M)(\omega_1r_1+\cdots+\omega_sr_s)}
\left(\sum\limits_{j=1}^p l_j-\sum\limits_{j \in \bbJ_k, k=1}^K l_j\right)
\label{ce}
\ee
is a reasonable estimate of the noise variance $\sigma^2$, where the 
$l_j$'s are the sample eigenvalues of $\gS$ and the $\bbJ_k$'s are the sets of ranks of the spikes.

As is well known, when the dimensionality $p$ is large compared to the sample size, the sample spiked eigenvalues do not converge to the population ones.
Thus, the estimator (\ref{ce}) has a negative bias. 
By the CLT proposed in Theorem~\ref{Th1},
we establish the following CLT for the estimator $\hat\sigma^2$
in the high-dimensional setting, which can be used to
identify the bias of estimator (\ref{ce}).

\begin{theorem}
\label{CLTsigma}
For the spiked model (\ref{51spikedmodel}) with the spectrum in (\ref{Spec}),
we assume that the assumptions in Theorem~\ref{Th1} hold and $c_n=p/n\to c > 0$ when both the dimensionality $p$ and sample size $n$ go to infinity.
Then  it is concluded that
\[
 \nu_x^{-\frac{1}{2}}\left\{
 (\hat\sigma^2-\sigma^2)(p-M)\sum\limits_{i=1}^s\omega_ir_i
 +b_(\tilde\alpha_k, \sigma^2)-\mu_x\right\}\Rightarrow \mathcal{N} \left( 0,
 1\right), 
\]
where 
 \begin{align*}
 b_(\tilde\alpha_k, \sigma^2)&=\sum\limits_{k=1}^K\sum\limits_{i=1}^s\frac{m_kc\tilde\alpha_k\sigma^2r_i\omega_i}{\tilde\alpha_k-r_i}, \non
   \mu_x&=\!-\frac{q}{2\pi i} \!\oint \frac{\!c\um^2(z)\big[c\um(z) \int t\{1+t\underline{m}(z)\}^{-1} \md H(t)-1\big]
 }{\left[1\!-\!c\!\int \underline{m}^2(z)t^2 \{1+t\underline{m}(z)\}^{-2}\md H(t)\right ]^2 }\non
 & \times \int t^2\{1+t\underline{m}(z)\}^{-3} \md H(t)\md z\non
  &-\frac{\beta c }{2 \pi i} \oint \um^2(z)\big[-1+c\um(z) \int t\{1+t\underline{m}(z)\}^{-1} \md H(t)\big]\non
&\times \frac{ \int t\{1+t\underline{m}(z)\}^{-1} \md H(t) \int \{1+t\underline{m}(z)\}^{-2} \md H(t)}
{1\!-\!c\!\int \underline{m}^2(z)t^2 \{1+t\underline{m}(z)\}^{-2}\md H(t)} \md z,\non
\nu_x&=-\frac{q+1}{4\pi^2}\oint\oint\frac{
\big[-\um^{-1}(z_1)+c \int t\{1+t\underline{m}(z_1)\}^{-1} \md H(t)\big]
}{\{\underline{m}(z_1)-\underline{m}(z_2)\}^2}\non
& \times \big[-\um^{-1}(z_2)+c \int t\{1+t\underline{m}(z_2)\}^{-1} \md H(t)\big]
\md\underline{m}(z_1)\md \underline{m}(z_2) \non
&-\frac{\beta c}{4\pi^2}\oint\oint \big[-\um^{-1}(z_1)+c \int t\{1+t\underline{m}(z_1)\}^{-1} \md H(t)\big]\non
& \times\big[-\um^{-1}(z_2)+c \int t\{1+t\underline{m}(z_2)\}^{-1} \md H(t)\big]\non
& \times \int \frac{t \md H(t)}{\{1+t\underline{m}(z_1)\}^2}\cdot \int \frac{t \md H(t)}{\{1+t\underline{m}(z_2)\}^2} ~\md\underline{m}(z_1)\md \underline{m}(z_2).
 \end{align*}
\end{theorem}
We note that the $b(\tilde\alpha_k, \sigma^2)$ depends on the number $m_k$, the LSD $H(t)$, and the values of the population spikes, which are most likely to be unknown in practice. 
To estimate these estimators, we refer the literature discussed in Section~\ref{Sec2}.

We now use the above theorem to correct the bias of $\hat \sigma^2$. As shown in Theorem~\ref{CLTsigma}, the bias of the estimator is also related to the unknown parameter $\sigma^2$, which will be estimated. Thus, a plug-in estimator is given in the following corollary. 
\begin{corollary}
\label{cbe}
For the spiked model (\ref{51spikedmodel}), as a result of Theorem~\ref{CLTsigma}, a plug-in estimator is given as
\be
\hat\sigma^2_{c}=\hat\sigma^2+\frac{b_(\tilde\alpha_k, \hat\sigma^2)-\mu_x}{(p-M)\sum\limits_{i=1}^s\omega_ir_i},\label{sigc2}
\ee
which is referred as the bias-corrected estimator.
\end{corollary}
Therefore, the asymptotic distribution of the bias-corrected estimator $\hat\sigma^2_{c}$ is a natural consequence of Theorem~\ref{CLTsigma}.

\begin{theorem}
If the conditions in Theorem~\ref{CLTsigma} are all satisfied, then the following conclusion naturally holds
\[
 \nu_x^{-\frac{1}{2}}(p-M)\sum\limits_{i=1}^s\omega_ir_i
 (\hat\sigma^2_c-\sigma^2)
\Rightarrow \mathcal{N} \left( 0,
 1\right).
\]
\end{theorem}

\subsection{Simulation study for estimation of noise variance}{\label{sec3.2}

For the estimation of the noise variance in the spiked model (\ref{51spikedmodel}), 
 we establish the following models:
\begin{description}
\item[Model~3]  Assume that $\gS=\sigma^2\Lambda$, where $\Lambda$ is defined in Model~1 and $\sigma^2=4$.
 \item[Model~4] Assume that $\gS=\sigma^2U_0 \Lambda U_0^*$, where $U_0$ and $\Lambda$ are defined in Model 2 and $\sigma^2=4$. 
\end{description} 
 Simulations were conducted to compare our proposed estimator $\hat\sigma^2_{c}$ in (\ref{sigc2}) with the other estimation methods, such as the maximum likelihood estimation (MLE) $\hat\sigma^2$ defined in (\ref{ce}), the estimator $\hat\sigma^2_{*}$ presented by \cite{Passemieretal2017}, the estimator $\hat\sigma^2_{us}$ presented by 
 \cite{UlfarssonSolo2008},
 and the estimator $\hat\sigma^2_m$ presented by 
 \cite{JohnstoneLu2009}. 
The Gaussian and Gamma population assumptions in Section~\ref{Sec2} were still used.
The mean absolute error (MAE) and mean squared error (MSE) of these estimators for Model~4 are reported in Table~\ref{tab:21} 
and \ref{tab:22} for 1000 replicates. 
Similar results for Model~3 are presented in the supplementary material.

\begin{table}
\caption{MAEs and MSEs of $\hat\sigma^2_c$ compared with those of existing estimators of the noise variance for Model~4 under Gaussian assumption.}
 \label{tab:21}\par
 \resizebox{\linewidth}{!}{\begin{tabular}{|lcccccc|} \hline
 \multicolumn{2}{|c}{Estimators} & $\hat\sigma^2_c$  & $\hat\sigma^2$ & $\hat\sigma^2_{*}$ &
 $\hat\sigma^2_{US}$ & $\hat\sigma^2_m$\\ \hline
       %      \multicolumn{7}{|c|}{Model~4 under Gaussian assumption}\\ \hline 
 $p=50;~ n=100$ &MAE& \bf 0.0672 &0.1091 &$ 6.87\times 10^2$ & 0.1195 &3.3 049\\[-0.5mm] 
               &MSE&  \bf 0.0071 &0.0166 & $ 6.51\times 10^5$ & 0.0220 &$11.103$\\[-0.5mm] 
 $p=100;~ n=200$ &MAE& \bf 0.0335 &0.0569 &$ 2.77\times 10^2$ & 0.0647 &1.7115\\[-0.5mm] 
               &MSE&  \bf 0.0018 &0.0045 & $ 1.17\times 10^5$ & 0.0065 &2.9538\\ [-0.5mm]
$p=200;~ n=400$ &MAE&  \bf 0.0159 &0.0277 &$ 8.51\times 10^1$ & 0.0335 &0.7907\\[-0.5mm] 
               &MSE& \bf 0.0004 & 0.0011 &$ 1.59\times 10^4$ & 0.0017 &0.6279\\ \hline
 $p=100;~ n=100$ &MAE& \bf 0.0549 &0.1154 &$ 1.09\times 10^3$ & 0.2239 &1.6165\\[-0.5mm] 
               &MSE&  \bf 0.0062 &0.0165 & $ 1.66\times 10^6$ & 0.0628 &2.6569\\[-0.5mm]               
 $p=200;~ n=200$ &MAE&  \bf 0.0257 &0.0551 &$ 5.13\times 10^2$ & 0.1102 &0.7743\\ [-0.5mm]
               &MSE& \bf 0.0010 &0.0038 &$ 3.65\times 10^5$ & 0.0156 &0.6047\\[-0.5mm]                
$p=400;~ n=400$ &MAE& \bf 0.0127 &0.0265 &$ 1.98\times 10^2$ & 0.0558 &0.4261\\ [-0.5mm]
               &MSE& \bf 0.0003 &0.0009 & $ 5.58\times 10^4$ & 0.0041 &0.1824\\ \hline
 $p=150;~ n=100$ &MAE& \bf 0.0398 &0.1137 &$ 2.02\times 10^3$ & 3.2501 &1.0087\\ [-0.5mm]
               &MSE&  \bf 0.0025 &0.0150 & $ 4.07\times 10^6$ & 10.564 &1.0320\\ [-0.5mm]
$p=300;~ n=200$ &MAE& \bf 0.0195 &0.0573 & $ 7.73\times 10^2$ & 3.2277 &0.5286\\ [-0.5mm]
               &MSE& \bf 0.0006 &0.0038 & $ 5.98\times 10^5$ & 10.418 &0.2816\\ [-0.5mm]
$p=600;~ n=400$ &MAE& \bf 0.0097 &0.0289 & $ 3.77\times 10^2$ & 3.2167 &0.2778\\ [-0.5mm]
               &MSE& \bf 0.0001 &0.0009 & $ 1.42\times 10^5$ & 10.347 &0.0776\\ \hline 
    \end{tabular}} 
\end{table}%   

\begin{table}
\caption{MAEs and MSEs of $\hat\sigma^2_c$ compared with those of existing estimators of the noise variance for Model~4 under Gamma assumption.}
 \label{tab:22}\par
 \resizebox{\linewidth}{!}{\begin{tabular}{|lcccccc|} \hline
 \multicolumn{2}{|c}{Estimators} & $\hat\sigma^2_c$  & $\hat\sigma^2$ & $\hat\sigma^2_{*}$ &
 $\hat\sigma^2_{US}$ & $\hat\sigma^2_m$\\ \hline                        
    %       \multicolumn{7}{|c|}{Model~4 under Gamma assumption}\\ \hline  
 $p=50;~ n=100$ &MAE& \bf 0.0834 &0.1226 &$ 6.93\times 10^2$ & 0.1407 &3.0216\\[-0.5mm] 
               &MSE&  \bf 0.0108 &0.0219 & $ 6.54\times 10^5$ & 0.0308 &9.3263\\ [-0.5mm]
 $p=100;~ n=200$ &MAE&  \bf 0.0426 &0.0625 & $ 2.44\times 10^2$ & 0.0727 &1.5157\\ [-0.5mm]
               &MSE& \bf 0.0028 &0.0057 & $ 9.48\times 10^4$ & 0.0083 &2.3219\\ [-0.5mm]
$p=200;~ n=400$ &MAE& \bf 0.0223 &0.0314 & $ 8.53\times 10^1$ & 0.0386 &0.8012\\ [-0.5mm]
               &MSE& \bf 0.0008 & 0.0014 &$ 1.59\times 10^4$ & 0.0023 &0.6454\\ \hline
 $p=100;~ n=100$ &MAE& \bf 0.0706 &0.1149 &$ 1.12\times 10^3$ & 0.2546 &1.5130\\ [-0.5mm]
               &MSE&  \bf 0.0102 &0.0178 & $ 1.63\times 10^6$ & 0.0796 &2.3255\\ [-0.5mm]
 $p=200;~ n=200$ &MAE&  \bf 0.0322 &0.0588 & $ 4.85\times 10^2$ & 0.1301 &0.7568\\[-0.5mm] 
               &MSE& \bf 0.0019 &0.0046 & $ 3.28\times 10^5$ & 0.0212 &0.5788\\   [-0.5mm]             
$p=400;~ n=400$ &MAE& \bf 0.0167 &0.0281 & $ 1.99\times 10^2$ & 0.0647 &0.4352\\[-0.5mm] 
               &MSE& \bf 0.0008 &0.0011 & $ 5.63\times 10^4$ & 0.0054 &0.1905\\ \hline
 $p=150;~ n=100$ &MAE& \bf 0.0511 &0.1183 &$ 2.02\times 10^3$ & 3.2568&1.0796\\[-0.5mm] 
               &MSE&  \bf 0.0041 &0.0175 & $ 4.06\times 10^6$ & 10.608&1.1852\\ [-0.5mm]
 $p=300;~ n=200$ &MAE& \bf 0.0263 &0.0576& $ 7.73\times 10^2$ & 3.2311 &0.5576\\ [-0.5mm]
               &MSE& \bf 0.0011 &0.0042 & $ 5.98\times 10^5$ & 10.439 &0.3144\\ [-0.5mm]
$p=600;~ n=400$ &MAE& \bf 0.0126 &0.0293 & $ 3.77\times 10^2$ & 3.2184 &0.2839\\ [-0.5mm]
               &MSE& \bf 0.0003 &0.0011 & $ 1.42\times 10^5$ & 10.358 &0.0813\\ \hline
  \end{tabular}} 
\end{table}%

As shown by the simulated results, our proposed method yielded the lowest MAEs and MSEs for different populations and models. Furthermore, the advantage of our estimation became increasingly evident as the dimensionality increased.
The estimator $\hat\sigma^2_{*}$ presented by \cite{Passemieretal2017} performed well for the  {diagonal} population covariance matrix with only extremely large spikes.
Their method involves a consistent estimate of $\tilde\alpha_k$ obtained by solving the equation 
$l_{j} \rightarrow \sigma^2 \{\tilde \alpha_k+{c\tilde \alpha_k}/{(\tilde \alpha_k-1)}\}$.
However, for the general spiked matrix with both extremely large and small spikes, their method always yielded estimates close to 1 for the the small spikes 0.1 and 0.2.
Thus, the estimates of $\sigma^2$ were completely ineffective.

\section{Application to testing equality of smallest roots in a probabilistic principal component
analysis model} \label{Sec4}

Suppose that the observable covariance matrix 
\be
\gS=\bA\bA'+\sigma^2 \bI,\label{52spikedmodel}
\ee
has a characteristic root of $\sigma^2$ with multiplicity $p-M$,
where 
 $\bA'\bA$ is the positive semidefinite matrix of rank $M$.
 We denote the population eigenvalues of $\gS$
 as $\lambda_1,\cdots, \lambda_p$ in descending order.
 We then consider testing the null hypothesis that 
 \be 
 \mathcal{H_0}: \lambda_{M+1}=\cdots= \lambda_p.\label{H0lq}
 \ee
This is equivalent to the null hypothesis that $\gS=\bA\bA'+\sigma^2 \bI$ when 
 $\bA'\bA$ is positive semidefinite of rank $M$.
 
 As shown in Section~11.7 in the publication of \cite{Anderson2003}, the pseudo-likelihood
 ratio criterion is the statistic $L$ defined in (\ref{Lstat}), where the $l_i$'s are the sample eigenvalues. Moreover, $-2\log L$ has a limiting $\chi^2$-distribution with 
 $(p-M+2)(p-M-1)/2$ degrees of freedom.
 However, this conclusion loses its effect when the dimension
 $p$ goes to infinity. Some previous researchers focused on this problem, such as \cite{Passemieretal2017}, who proposed a 
 goodness-of-fit test for a probabilistic principal component
analysis model with the form of (\ref{52spikedmodel}). However, their result only applied to a Gaussian population. Therefore,
 we propose a corrected test statistic  $-2\log L/\{n(p-M)\}$ and derive its limiting distribution by Theorem~\ref{Th1},
 which is widely used without a Gaussian assumption constraint.

 \begin{theorem}
 For the test problem (\ref{H0lq}), we suppose that the standardized entries for the model (\ref{52spikedmodel}) satisfy condition (\ref{tail}) and $p/n = c_n \to c >0$ when both $n$ and $p$ go to infinity simultaneously.
For the test statistic $ -{2\log L}\big/\{n(p-M)\}$, it is determined that 
\[
T_L= \nu_L^{-\frac{1}{2}}\left\{ -\frac{2\log L}{n(p-M)}-\log \left(\frac{b_{x,\sigma^2}}{p-M}\right)+\frac{b_{\log,\sigma^2}+\mu_{\log, \sigma^2}}{p-M}\right\}\Rightarrow \mathcal{N} \left( 0,
 1\right), \label{CLTL}
\]
 where $\nu_L$ is expressed by (\ref{varL})
 and $b_{x,\sigma^2}$, $b_{\log,\sigma^2}$, $\mu_{\log,\sigma^2}$, $\nu_{x,\sigma^2}$ and $\nu_{\log,\sigma^2}$ are defined in (\ref{bmunu}).
   \label{Th3}
 \end{theorem}
\proof 

 We set $\alpha_k^*, k =1,\ldots,K$ as the $M$ non-zero eigenvalues of $\bA\bA'$ with $m_k$ multiplicity. The spikes of $\gS$
 in model (\ref{52spikedmodel}) are $\sigma^2\tilde \alpha_k, k=1, \ldots, K$, where $\tilde \alpha_k=\alpha_k^*/\sigma^2+1$ and also has multiplicity $m_k$. We define 
$
\Delta_1=\sum_{i=M+1}^p l_i-b_{x,\sigma^2}-\mu_{x,\sigma^2}$ and  $\Delta_2=\sum_{i=M+1}^p \log l_i-b_{\log,\sigma^2}-\mu_{\log, \sigma^2}.
$
By Theorem ~\ref{Th1}, we can determine that 
 \be
T_{x, \sigma^2}=\nu_{x,\sigma^2}^{-\frac{1}{2}} \Delta_1\rightarrow \mathcal{N}(0,1) ~~\text{and }~~ T_{\log, \sigma^2}=\nu_{\log, \sigma^2}^{-\frac{1}{2}} \Delta_2\rightarrow \mathcal{N}(0,1),\label{twonormal}
\ee
where 
\bqa
b_{x,\sigma^2}&=&(p-M)\sigma^2-\sum\limits_{i=1}^K \frac{m_kc\sigma^2\tilde \alpha_k}{\tilde \alpha_k-1},~
\mu_{x,\sigma^2}=0,~\nu_{x,\sigma^2}=(q+1+\beta)c\sigma^4,\non
b_{\log,\sigma^2}&=&p\left\{\frac{(c-1)}{c}\log(1-c)-1\right\}-\sum\limits_{i=1}^K m_k \log (1+\frac{c}{ \alpha_k-1}), \non
\mu_{\log,\sigma^2}&=&\frac{q}{2}\log (1-c)-\frac{1}{2}\beta c, \quad \nu_{\log,\sigma^2}=-(q+1)\log (1-c)+\beta c.
\label{bmunu}
\eqa
 %Here, $\beta={\rE|x_{11}|^4-q-2}$, with $q=1$ for the real case and $q=0$ for the complex case.
 The detailed calculations are similar to Examples~1 and 2. 
Furthermore, 
by the expression of $L$ given in (\ref{Lstat}) and the Taylor expansion, it follows that
 \bqa
 &&-\frac{2\log L}{n(p-M)}=\log(\sum\limits_{i=M+1}^p l_i)-\frac{1}{p-M}\sum\limits_{i=M+1}^p \log l_i-\log(p-M)\non
 &=& \log(\Delta_1+b_{x,\sigma^2}+\mu_x) -\frac{1}{p-M}(\Delta_2+b_{\log}+\mu_{\log})-\log(p-M)\non
 &=& \log\left(\!\frac{b_{x,\sigma^2}+\mu_{x,\sigma^2}}{p-M}\!\right)\!+\!\frac{\Delta_1}{b_{x,\sigma^2}+\mu_{x,\sigma^2}}\!-\!\frac{\Delta_2}{p-M}-\frac{b_{\log,\sigma^2}\!+\!\mu_{\log,\sigma^2}}{p-M}.\label{Ld1d2}
 \eqa
 Based on equation (\ref{twonormal}), the following is obtained:
\be
\frac{\Delta_1}{b_{x,\sigma^2}+\mu_x}-\frac{\Delta_2}{p-M}\rightarrow \mathcal{N}(0, \nu_L) ,\label{d1d2}
\ee
where 
\be
\nu_L=\frac{\nu_{x,\sigma^2}(p-M-2b_{x,\sigma^2})}{(p-M)b_{x,\sigma^2}^2}+\frac{\nu_{\log,\sigma^2}}{(p-M)^2}\label{varL}
\ee
is calculated according to (\ref{bmunu}) and (\ref{d1d2}).
 
Thus, by (\ref{Ld1d2}) and (\ref{d1d2}), 
the proof is complete. 
 
\subsection{Simulation study for testing equality of smallest roots}

Simulations were conducted to compare our proposed test statistic $T_L$ with the classical pseudo-likelihood ratio test statistic ($T_{PLR}$) and the corrected likelihood
ratio test (CLRT) presented by \cite{Passemieretal2017}. These test methods all rely on the pseudo-likelihood function and therefore have good statistical properties, but they can only be used for the case of $p<n$, 
 due to their correlation with
 the log function. To expand the application of our method, the test statistic $T_{x,\sigma^2}$ in (\ref{twonormal}) was also used as a supplement.
The value of the test statistic of the CLRT presented by \cite{Passemieretal2017} cannot be calculated
under our general model assumptions, as mentioned in Section~{\ref{sec3.2}}.
We used $\text{CLRT}_r$ to represent the test method of \cite{Passemieretal2017} with their estimated $\tilde\alpha_k$ replaced by the real values of $\tilde\alpha_k$.
Models 1 and 2 and the population assumptions in Section 2 were again used here. 
 The empirical sizes of the competitive tests for hypothesis (\ref{H0lq}) were calculated for 1000 replicates.
 The simulated results for Model 2 are presented 
 in the Figures \ref{fig:tu2} and \ref{fig:tu4}.
 Similar figures for Model~1 are included in the supplementary material.
 \begin{figure}[t!]
\includegraphics[width = .48\textwidth] {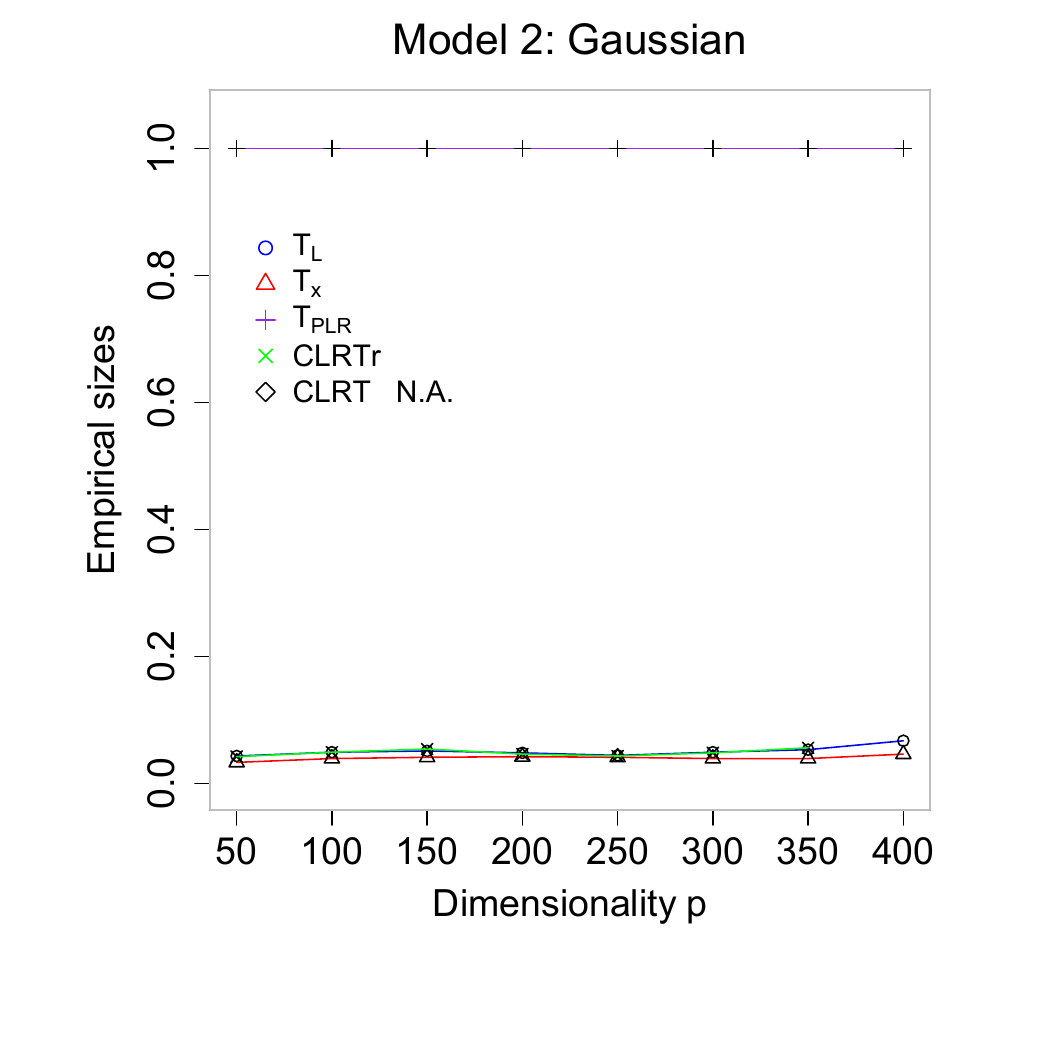}
\quad\includegraphics[width = .48\textwidth] {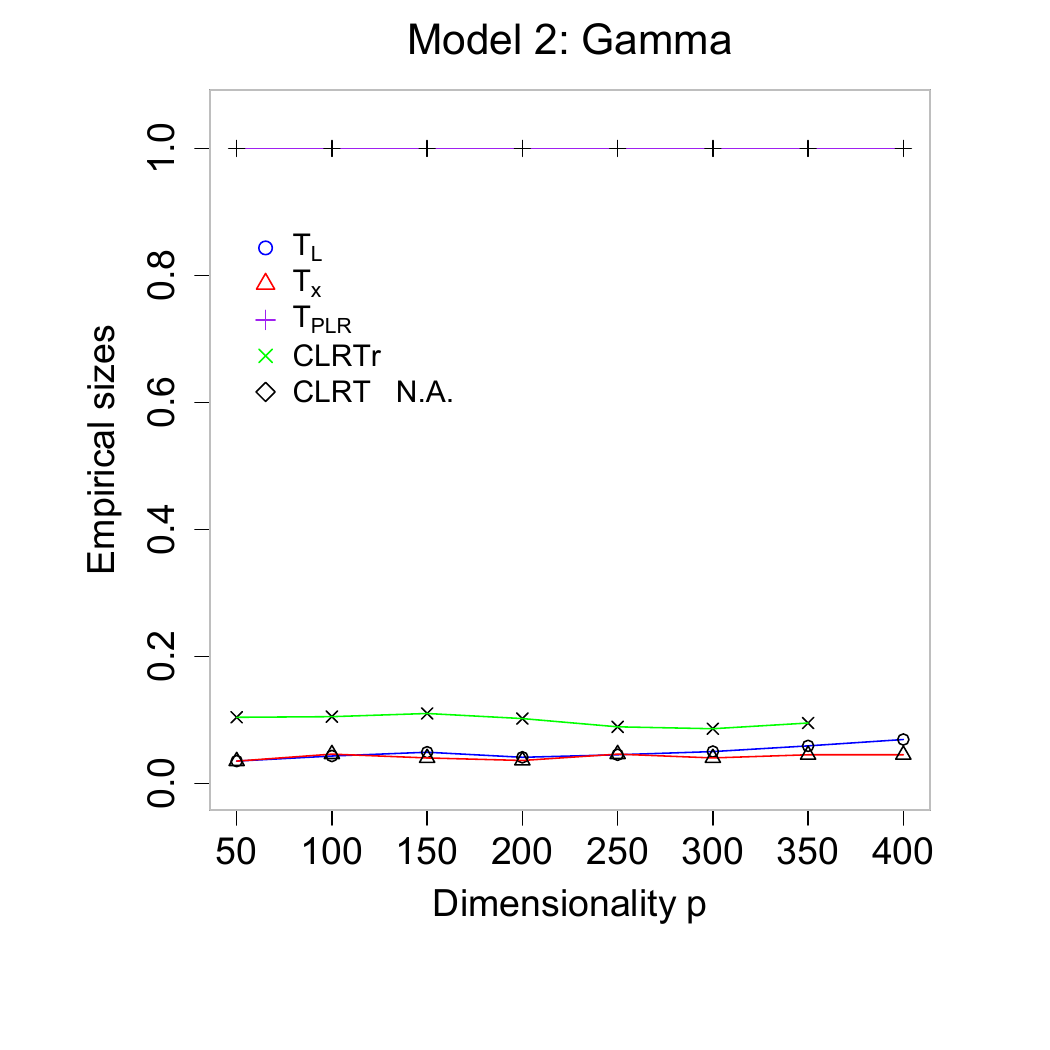}
\caption{Empirical sizes of the competitive tests for hypothesis (\ref{H0lq}) when $n=500$ and $p/n<1$.
 }\label{fig:tu2}
\end{figure}

 \begin{figure}[t!]
\includegraphics[width = .48\textwidth] {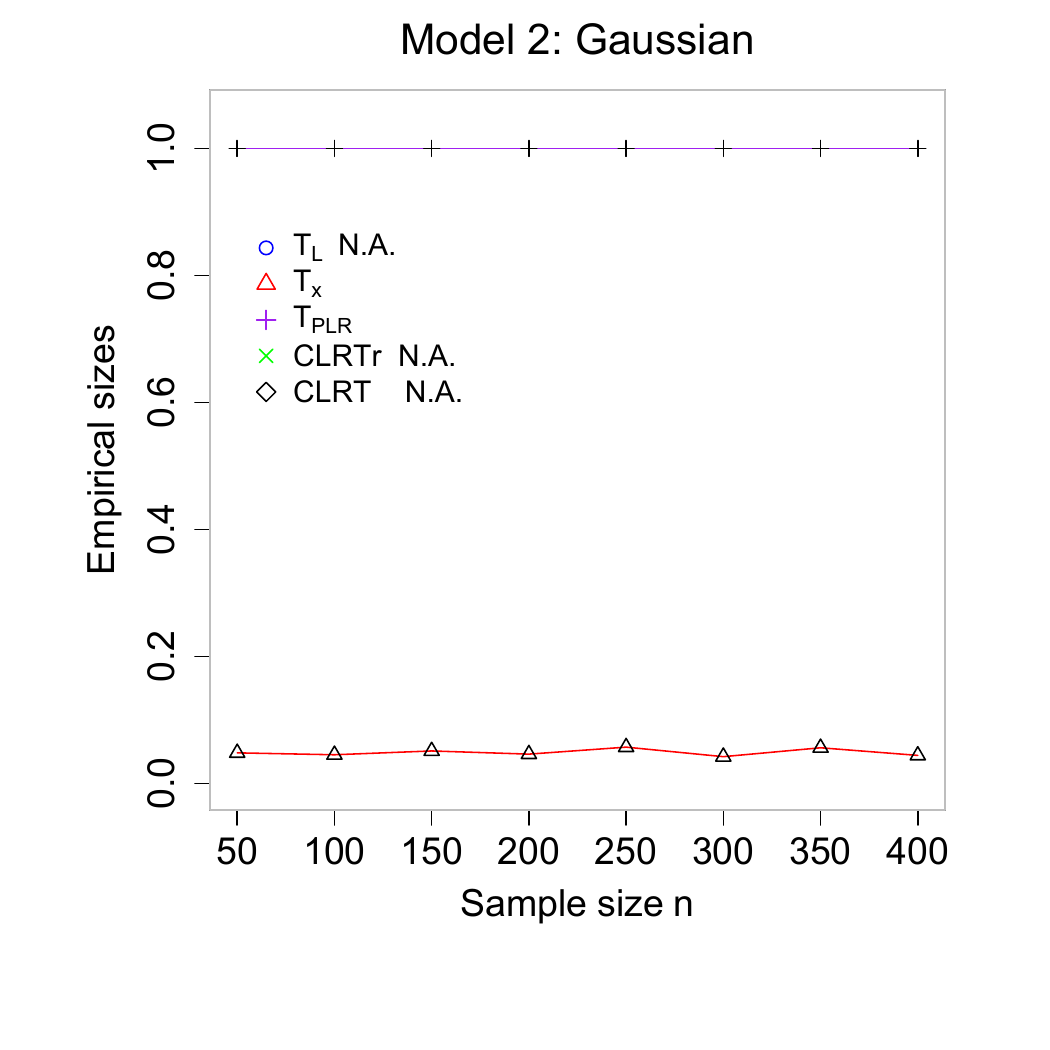}
\quad\includegraphics[width = .48\textwidth] {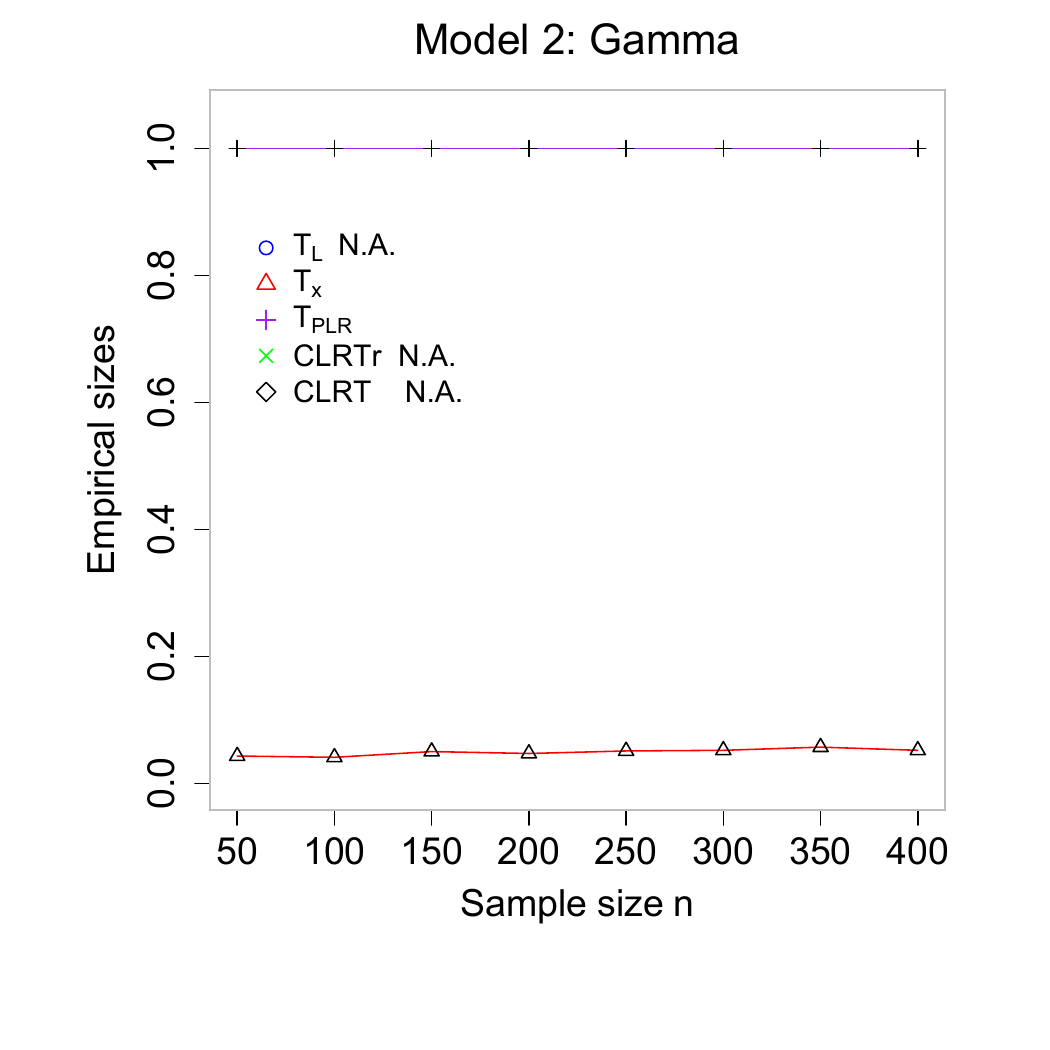}
\caption{Empirical sizes of the competitive tests for hypothesis (\ref{H0lq}) when $p=500$ and $p/n>1$.
 }\label{fig:tu4}
\end{figure}

Our proposed test statistics $T_L$ and $T_x$ provide the empirical sizes around the selected test level of 5\%. 
Moreover, $T_x$ can be applied more broadly to the case with $p>n$. 
Meanwhile, the classical test statistic $T_{PLR}$ absolutely rejected whatever the null hypothesis was when the dimensionality increased. 
As explained in Section~{\ref{sec3.2}}, the CLRT is not available (N.A.)
under our model assumptions, which has both extremely large and small spikes. 
Even if we used the true values of the spikes instead of their estimates in the CLRT, there were still problems. First, in the case of the Gamma population, the empirical size of $\text{CLRT}_r$ was significantly higher than the given test level. 
Second, for the high-dimension case $p=400, n=500$, the value of the test statistic of $\text{CLRT}_r$ still could not be calculated.

 %%%%%%%%%%%%%%%%%%%%%%%%%%%%%%%%%%%%%%%%%%%%%%%%
 %real data
%%%%%%%%%%%%%%%%%%%%%%%%%%%%%%%%%%%%%%%%%%%%%%%%%%

\section{Real Data Analysis}\label{Sec5}
  
To demonstrate the feasibility of our proposed test method, 
  we used two real datasets. 
  The first dataset
was the environmental dataset for world countries, which is freely available from the website  
\url{https://www.kaggle.com/zanderventer/environmental-variables-for-world-countries}. Since country-level social and economic statistics are often limited to socio-economic data, this dataset enables people to use 
 environmental statistics to predict social and economic data. Determining how many environmental variables have a significant impact on socioeconomic status is an important problem.
 The dataset 
 consists of 243 countries and 27 environmental variables. We applied our testing method to determine the number of the spikes of the covariance matrix generated from the standardized data.
 We used 188 observations of these environmental variables without missing values. The  $p$-values of the sequential tests are listed in Table~\ref{tr1}. 
We inferred that 
the true value of the number of spikes appeared at the first inflection point of the $p$-value, where the first local maximum value occurred. In this way, we could determine the number of spikes and provide appropriate estimates.
 The estimates of the values of the spikes are also included. 
 
 The second dataset considered was the {Wisconsin Breast Cancer Diagnosis dataset}, which was 
 downloaded from \url{
 https://archive.ics.uci.edu/ml/datasets/Breast+Cancer+Wisconsin+\%28Diagnostic\%29}.
This dataset contains 569 instances, with 357 benign (62.7\%) and 212 (37.3\%) malignant cases of breast cancer.
  Thirty real-valued input features were computed for each cell nucleus, as well as two nominal features: ID number and diagnosis. 
 The database adopted here was a standardized version of the original Wisconsin Breast Cancer Diagnosis dataset. Our proposed 
 testing method to determine the number of spikes
 was applied for all the instances.
  The test results are listed in Table~\ref{tr2}.
  
  \begin{table}[H]
\caption{Estimates of the number and values of population spikes for the environmental dataset by country.}
 \label{tr1}\par
 \resizebox{\linewidth}{!}{\begin{tabular}{|ccccccccc|} \hline
 Values of $M_0$: &$ 1 $&$ 2 $& $ 3 $& $4$& $5$& $6$& $7$&\\[1mm]
$p$-values:& 0 & 0.0211 & 0.3498 & 0.6976 & \bf 0.9488& 0.4211 & 02205 &\\[1mm]
Estimated number: & \multicolumn{8}{c|} {\bf 5} \\[1mm]
{Estimated values of spikes:} &&$\hat\alpha_1$ & $\hat\alpha_2$& $\hat\alpha_3 $ &$\hat\alpha_4$& $\hat\alpha_5$&&\\
&& 9.7848 & 7.0903 & 2.1086 & 1.7522& 1.3634&& \\ \hline
  \end{tabular}} 
\end{table}%

   \begin{table}[H]
\caption{Estimates of the number and values of population spikes for the Wisconsin Breast Cancer Diagnosis dataset.}
 \label{tr2}\par
 \resizebox{\linewidth}{!}{\begin{tabular}{|cccccccc|} \hline
Values of $M_0$: &$ 1 $&$ 2 $& $ 3 $& $4$& $5$& $6$& \\[1mm]
$p$-values:& 0 &0 & $8.11\times 10^{-10}$& \bf 0.1026 & $1.18\times 10^{-14}$& 0 & \\[1mm]
Estimated number: & \multicolumn{7}{c|} {\bf 4} \\[1mm]
{Estimated values of spikes:} &&$\hat\alpha_1$ & $\hat\alpha_2$& $\hat\alpha_3 $ &$\hat\alpha_4$& &\\
&& 13.1817 & 5.6174 & 2.7219 & 1.9264& & \\ \hline
  \end{tabular}} 
\end{table}%
 
%%%%%%%%%%%%%%%%%%%%%%%%%%%%%%%%%%%%%%%%%%%%%%%%%%%
%  Conclusion
%%%%%%%%%%%%%%%%%%%%%%%%%%%%%%%%%%%%%%%%%%%%%%%%%%%

\section{Conclusion}\label{Sec6}

In this paper, we established a universal test for the number of spikes in a high-dimensional generalized spiked model with more relaxed assumptions than previous tests.
The conclusion was applied to two typical statistical problems, and the effectiveness of our proposed test method was proven by simulation results.
This paper only focused on the one-sample spiked model related to the covariance matrix. We will continue to study the two-sample spiked model involved with the Fisher-matrix in future work.

%\section*{Supplementary Materials}
%
% The supplementary material  for ``A universal test on  spikes in a high-dimensional generalized spiked model and its applications''
%  is available online and includes some simulation results as well as detailed proofs for Examples~\ref{eg:1} and \ref{eg:2}.
%\par
%%%%%%%%%%%%%%%%%%%%%%%%%%%%%%%%%%%%%%%%%%%%%%%%%%%%%%%%%%%%%%%%%%%%%%%%%%%%%%%%%%%%%%%%%%%%%%%%%%%%%%%%%%%%%%%%%%%%%%%%%%%%
\section*{Acknowledgments}
%We are grateful to the Editor, the Associate Editors and referees for their constructive and helpful comments. 
We thank LetPub (www.letpub.com) for linguistic assistance and pre-submission expert review.
The author was supported by NSFC Grant 11971371 and  Natural Science Foundation of Shaanxi Province 2020JM-049.

\par

\bibhang=1.7pc
\bibsep=2pt
\fontsize{9}{14pt plus.8pt minus .6pt}\selectfont
\renewcommand\bibname{\large \bf References}
%\begin{thebibliography}{11}
\expandafter\ifx\csname
natexlab\endcsname\relax\def\natexlab#1{#1}\fi
\expandafter\ifx\csname url\endcsname\relax
  \def\url#1{\texttt{#1}}\fi
\expandafter\ifx\csname urlprefix\endcsname\relax\def\urlprefix{URL}\fi

%% use bibfile 
%  \bibliographystyle{chicago}      % Chicago style, author-year citations
%  \bibliography{bibfile}   % name your BibTeX data base

%%  Another method

%%%%%%%%%%%%%%%%%%%%%%%%%%%%%%%%%%%%%%%%%%%%%%%%%%%%%%%%%%%%%%%%%%%%%%%%%%%%%%%%%%%%%%%%%%%%%%%%%%%%%%%%%%%%%%%%%%%%%%%%%%%%
\vskip .65cm
\noindent
School of Mathematics and Statistics, Xi'an Jiaotong University,
No.28, Xianning West Road, Xi'an, Shannxi, 710049, P.R. China.
\vskip 2pt
\noindent
E-mail: (jiangdd@xjtu.edu.cn)
\vskip 2pt

% \vskip .3cm
%\centerline{(Received ???? 20??; accepted ???? 20??)}\par
\end{document}